\documentclass[aps,prb,twocolumn,superscriptaddress,showpacs]{revtex4}

\usepackage{epsfig}
\usepackage{lscape}
\usepackage{amsmath}

\begin{document}

\title{Atomic and molecular adsorption on transition-metal carbide
  (111) surfaces from density-functional theory: A trend study of
  surface electronic factors}    

\author{A. Vojvodic}
\affiliation{Materials and Surface Theory Group, Department of Applied
  Physics, Chalmers University of Technology, SE-412 96 G\"{o}teborg,
  Sweden}
\email{alevoj@chalmers.se} 

\author{C. Ruberto}
\affiliation{Materials and Surface Theory Group, Department of Applied
  Physics, Chalmers University of Technology, SE-412 96 G\"{o}teborg,
  Sweden} 

\author{B.~I. Lundqvist}
\affiliation{Materials and Surface Theory Group, Department of Applied
  Physics, Chalmers University of Technology, SE-412 96 G\"{o}teborg,
  Sweden} 
\affiliation{Center for Atomic-scale Materials Design, Department of
  Physics, Technical University of Denmark, DK-2800 Kongens Lyngby,
  Denmark}



\begin{abstract}
This study explores atomic and molecular adsorption on a number of
early transition-metal carbides (TMC's) by means of density-functional
theory calculations. The investigated substrates are the TM-terminated
TMC($111$) surfaces, of interest because of the presence of different
types of surface resonances (SR's) on them and because of their
technological importance in growth processes. Also, TM compounds have
shown potential in catalysis applications. Trend studies are conducted
with respect to both period and group in the periodic table, choosing
the substrates ScC, TiC, VC, ZrC, NbC, $\delta$-MoC, TaC, and WC (in
NaCl structure) and the adsorbates H, B, C, N, O, F, NH, NH$_2$, and
NH$_3$. Trends in adsorption strength are explained in terms of
surface electronic factors, by correlating the calculated adsorption
energy values with the calculated surface electronic structures. The
results are rationalized with use of a concerted-coupling model (CCM),
which has previously been applied succesfully to the description of
adsorption on TiC($111$) and TiN($111$) surfaces [Solid State
  Commun. \textbf{141}, 48 (2007)]. First, the clean TMC($111$)
surfaces are characterized by calculating surface energies, surface
relaxations, Bader charges, and surface-localized densities of states
(DOS's). Detailed comparisons between surface and bulk DOS's reveal
the existence of transition-metal localized SR's (TMSR's) in the
pseudogap and of several C-localized SR's (CSR's) in the upper valence
band on all considered TMC($111$) surfaces. The spatial extent and the
dangling bond nature of these SR's are supported by real-space
analyses of the calculated Kohn-Sham wave functions. Then, atomic and
molecular adsorption energies, geometries, and charge transfers are
presented. An analysis of the adsorbate-induced changes in surface
DOS's reveals a presence of both adsorbate--TMSR and adsorbate--CSR's
interactions, of varying strengths depending on the surface and the
adsorbate. These variations are correlated to the variations in
adsorption energies. The results are used to generalize the content
and applications of the previously proposed CCM to this larger class
of substrates and adsorbates. Implications for other classes of
materials, for catalysis, and for other surface processes are
discussed.   
\end{abstract}

\maketitle


\section{Introduction}

Many studies of TMX's (TM = transition metal and X = C or N) are
motivated by a curiosity on the properties of TMX's, such as their
mixture of covalency, ionicity, and metallicity, and by a suggested
importance for heterogeneous catalysis. The attention to TMX's as
potential catalysts started with an observation by Levy and Boudart
that the TMC's show a Pt-like behavior in several catalytic
reactions.\cite{LeBo73} According to more recent investigations,
``early TMC's and TMN's often demonstrate catalytic advantages over
their parent metals in activity, selectivity and resistance to
poisoning'' and ``for several reactions, such as hydrogenation
reactions, catalytic activities of TMC's and TMN's are approaching or
surpassing those of group VIII noble metals''.\cite{Chen} The TMX
surfaces are also technologically important as substrate materials in
growth processes, \textit{e.g.}, in wear-resistant multilayer coatings
on industrial cutting tools \cite{HaVu97,PrPfSa98} and for growth of
carbidic nanostructures.\cite{ItOsIcAi91,GuBaOtSo01} 

For catalytic applications, the stable or ideal surfaces are not
necessarily the most suitable ones. Often the best site for a reaction
is found on a stepped or in some way non-perfect surface,
\textit{e.g.}, at kinks or around defects, where less stable faces of
the material are exposed. Such sites often host surface states or
surface resonances (SR's). This calls for studies on surfaces that
present such surface states or resonances.   

In previous studies, the reactivities of the TiC($111$) and TiN($111$)
surfaces are attributed to the presence of SR's of both Ti and C/N
character.\cite{RuLu07,VoRuLu06,RuVoLu06,RuVoLu07} Calculated trends
in atomic adsorption strength are explained with a concerted-coupling
model (CCM), in which the atomic frontier orbitals interact with both
types of SR's. More recently, we have indentified a descriptor,
defined as the mean energy of the TM-derived SR (TMSR), for atomic and
molecular adsorption and for activation-energy
barriers.\cite{VoHeRuLu09} Hence, the existence of several linear
relations between the atomic and molecular adsorption energies,
``scaling relation'', have been shown.\cite{VoHeRuLu09} Such relations
are of importance in the design of novel types of
catalysts.~\cite{Abild-Pedersen07,Sehested2007,Studt,Norskov2009}     

This paper is devoted to a deeper and more generalized understanding
of the chemisorption on the TMC($111$) surfaces, thus also laying the
ground for the above mentioned descriptor. We extend the work done in
our previous study of the atomic adsorption on TiC($111$) and 
TiN($111$) and use density-functional theory (DFT) to investigate
whether the CCM is applicable to other TMC's. Our method consists of a
detailed study of the trends in reactivity along periods and groups of
the substrate parent metal and of the adsorbate, correlated with a
careful mapping and analysis of the underlying details of the changes
in surface electronic structure upon adsorption.   

\begin{figure}[h]
\centering
\includegraphics[width=0.25\textwidth]{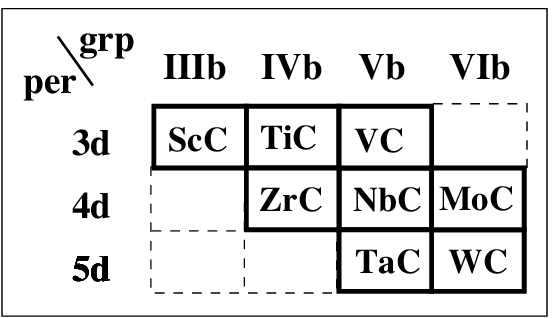}
\caption{\label{fig:def_TMC}
The early transition-metal carbides under investigation.} 
\end{figure}

The substrates chosen in our study are the TMC's formed with the
parent metals Sc, Ti, V, Zr, Nb, Mo, Ta, and W in NaCl structure (see
Fig.~\ref{fig:def_TMC}). As adsorbates we choose atomic H, B, C, N, O,
and F as well as the molecules NH, NH$_2$ and NH$_3$. These particular
choices of TMC's and adsorbates allow us to capture changes in
adsorption properties along both periods and groups in the periodic
table. All studied TMC's adopt a NaCl structure either in stable or in
metastable phase.   

%
%

Experiments show that after heating and ion or electron bombardment
under low-temperature and low-pressure conditions the ($111$) surfaces
of TiC,\cite{AonoTiC} VC,\cite{RundgrenVC} ZrC,\cite{HwangZrC}
NbC,\cite{EdamotoNbCLEED,HayamiNbC} and
TaC\cite{SoudaTaCLEED,HulbertTaC} are unreconstructed and TM
terminated. Therefore, our study deals with unreconstructed and
TM-terminated ($111$) TMC surfaces. Surface electronic structure
characterizations with angle-resolved photoemission studies (ARPES)
show the presence of surface states on the ($111$) surfaces of
TiC,~\cite{WeaverTiC,ZaimaTiC,EdamotoTiC} ZrC,~\cite{EdamotoZrC}
NbC,\cite{EdamotoNbC} and TaC.\cite{AnazawaTaC} Also, experiments show
that both H$_2$ and O$_2$ dissociate on
TiC($111$),\cite{ZaimaTiC,BrandshawTiC,SoudaTiC,EdamotoTiCH,EdamotoTiCO}
ZrC($111$),\cite{TokumitsuZrCH,NakaneZrCH,NodaZrCO} and 
NbC($111$).\cite{KojimaNbCO,EdamotoNbCO,HayamiNbCO,AizawaNbCO} Despite
this large number of experimental studies, theoretical investigations
on TMC($111$) surfaces are
scarce.\cite{TanTiC111,ZhangNbC111,Kitchin,ZhLiDiChZh02}     

This paper presents our results in the following order. First, a trend
study of the clean TM-terminated TMC($111$) surfaces is presented in
Section~\ref{sec:surface}. In particular we address the properties of
the surface electronic structures that are relevant for the adsorption
mechanisms. Then, in Section~\ref{sec:ads} the adsorption-energy
trends are presented together with the trends in the adorption-induced
changes in surface electronic structure. The different trends in
adsorption-energy and electronic-structure changes are analyzed and
discussed in Section~\ref{sec:discussion}, thus connecting our results
to the previously proposed CCM. The paper is concluded by
Section~\ref{sec:conclusion}, which summarizes the main conclusions of
our investigation and makes some outlooks to possible ramifications
and further investigations.


\section{Trends in surface properties}\label{sec:surface}

In this Section the computational details and results from our DFT
calculations on the clean TM-terminated TMC($111$) surfaces are 
presented. First, the stability of the ($111$) surfaces is considered 
by comparing the cleavage energies of these surfaces with the
corresponding results for the ($100$) surfaces. Then, the relaxed
surface structures are presented and compared with existing
results. This is followed by a charge transfer analysis and a detailed 
analysis of both the energy- and space-resolved surface densities of
states (DOS's). Particular emphasis is put on the presence and
character of SR's on the TMC($111$) surfaces. Throughout the
presentation, analyses of the trends with respect to the TM component
of the studied TMC's are made.

\subsection{Computational details}
\label{sec:surf_comp}

The surface calculations presented in this paper are performed within
the DFT formalism using the plane-wave pseudopotential code
Dacapo.\cite{dacapo} The ion-electron interaction is treated with
Vanderbilt ultrasoft pseudopotentials.\cite{Vanderbilt} The
exchange-correlation energy is included by the generalized gradient
approximation (GGA) using the PW91 functional.\cite{BuPeWa98} We
utilize a slab geometry, with slabs of $4$ to $8$ bilayers (a bilayer
being a unit of one TM layer and one C layer), a vacuum region
thickness corresponding to $5$ bilayers, that is, at least
$10.8$~{\AA}, and periodic boundary conditions. Each atomic layer is
composed of one atom in a ($1 \times 1$) geometry. The atoms in the
three (four) outermost atomic layers on the TM-terminated side of a
$4$ ($5-8$) bilayer thick slab are allowed to relax until the sum of
the remaining forces on all relaxed atoms is less than $0.05$~eV/\AA,
while the remaining atomic layers are fixed at the bulk geometry. A
Monkhorst-Pack sampling~\cite{MoPa76} of $8\times8\times1$ special
$k$-points and a plane-wave energy cutoff of $400$~eV are used. The
slab used to model the ($111$) surface is asymmetric, which gives rise
to a discontinuity in the electrostatic potential at the cell
boundary. This is corrected for by using the scheme in
Ref.~\onlinecite{Bengtsson}.     
 
To characterize the surfaces we utilize several electronic structure
tools. A Bader analysis is used to calculate the charge localization
around individual atoms.\cite{Bader, Baderalgorithm} Total and local,
that is, atom-projected, DOS's for the surface bilayer are obtained by
projecting the Kohn-Sham wave functions onto individual atomic
orbitals and plotted as a function of energy (relative to the Fermi
level $E_F$). To identify the surface specific properties that arise
upon creation of the surface, the surface DOS's are compared to the
bulk DOS's by studying the differences between the two
quantities. Information about the spatial localization of the
surface-localized states is provided by analyzing the space-resolved
surface DOS, that is, the Kohn-Sham wave functions.

\subsection{Cleavage energies}

The cleavage energy $E_{\text{cleav}}$, that is, the energy needed to
create two surfaces upon cleavage of a bulk structure along a specific
crystallographic plane, is calculated as  
\begin{equation}
E_{\text{cleav}} = E_{\text{slab}}(n) - n E_{\text{bulk}},
\end{equation}
where $E_{\text{slab}}(n)$ is the total energy of a slab with $n$ TMC
bilayers that exposes the two surfaces under investigation and    
\begin{equation}
E_{\text{bulk}} = E_{\text{slab}} (n) - E_{\text{slab}} (n-1)
\end{equation}
is the bulk energy of one TMC bilayer, if $n$ is sufficiently
large. Thus, the cleavage energy corresponds to the sum of the surface
energies of the two surfaces obtained upon cleavage. In the case of
calculations on stoichiometric TMC($111$) slabs, the cleavage energy
is equal to the sum of the surface energies of the TM-terminated
surface and of the C-terminated surface. 

Our calculated $E_{\text{cleav}}$ values for the TMC($111$) surfaces,
after relaxation of only the TM-terminated side of the slabs, are
given in Table~\ref{tab:TMC_surf_relax}. Along a period
$E_{\text{cleav}}$ shows a maximum for group IV, thus showing the same
variations as the ones found for the bulk cohesive energies in our
previous study.\cite{VoRu09} Down a group, the variations in
$E_{\text{cleav}}$ are small but discernible and do not show any
apparent correspondence to the bulk cohesive energies of
Ref.~\onlinecite{VoRu09}.    

According to the calculated $E_{\text{cleav}}$ values, TiC($111$) is
the surface that requires the most energy to create among the
considered TMC surfaces. Nevertheless, this surface is routinely grown
by chemical-vapor deposition (CVD) under high temperatures as
wear-resistant coating on industrial cutting tools.\cite{HaVu97,PrPfSa98}
The calculated $E_{\text{cleav}}$ value for the TiC($111$) surface
agrees well with those of previous DFT calculations.\cite{Dudiy,RuLu07}          

Compared to the TMC($100$) surfaces, the TMC($111$) surfaces are found
to have a higher $E_{\text{cleav}}$ value (see
Table~\ref{tab:TMC_surf_relax}) and are thus less stable. Also, the
variations in $E_{\text{cleav}}$ are larger for TMC($111$) than for
TMC($100$) surfaces. Both these properties can be attributed to the
polar nature of the ($111$) surface.\cite{Ta79}

\subsection{Surface geometry}
%
\begin{table*}
\caption{\label{tab:TMC_surf_relax}
  Relaxed surface energetics and geometry of the considered TMC's:
  Calculated cleavage energies $E_{\text{cleav}}$, in J/m$^2$, for the
  TM-terminated ($111$) and for the ($100$) surfaces; Perpendicular
  equilibrium distances $d_{ij}$, in absolute values (\AA ) and
  relative to the bulk values (in parentheses), between atomic layers
  at the TM-terminated ($111$) surfaces (index $1$: topmost surface
  layer; index $2$: second surface layer, \textit{etc.}). The bulk
  interlayer distances $\Delta$ and the available experimental surface
  interlayer distances $d_{12}^{\text{exp}}$ are also given. The
  calculated results are extracted from the $8$-bilayer slabs and
  differ by less than $1\%$ from the ones from the $4$-bilayer slabs.}            
\begin{center}
\begin{tabular}{ccccccccccc}
\hline
\hline
Period & Group& Surface& $E_{\text{cleav}}^{111}$ & $E_{\text{cleav}}^{100}$ & d$_{12}$ & d$_{23}$ & d$_{34}$ & d$_{45}$ & $\Delta$ & d$_{12}^{\text{exp}}$\\ 
\hline
    & III & ScC & $6.06$  & $-$    & $0.967 \ (-28.5\%)$ & $1.558 \ (15.2\%)$ & $1.303 \ (-3.6\%)$ & $1.375 \ (1.7\%)$ & $1.352$ & $-$\\
$3d$& IV  & TiC & $11.29$ & $3.29$\footnotemark[1] & $1.019 \ (-18.5\%)$ & $1.394 \ (11.5\%)$ & $1.196 \ (-4.4\%)$ & $1.278 \ (2.3\%)$ & $1.251$ & $0.87 \ (-30\%)$\footnotemark[2]\\      
    & V   & VC  & $9.14$  & $3.21$\footnotemark[1] & $0.961 \ (-19.8\%)$ & $1.293 \ (7.8\%)$ & $1.176 \ (-2.0\%)$ & $1.204 \ (0.3\%)$ & $1.202$ & $1.09 \ (-10\%)$\footnotemark[3]\\
\hline
    & IV  & ZrC & $10.09$ & $3.10$\footnotemark[1] & $1.117 \ (-17.8\%)$ & $1.513 \ (11.5\%)$ & $1.299 \ (-4.3\%)$ & $1.378 \ (1.5\%)$ & $1.357$ & $ -$\\ 
$4d$& V   & NbC & $8.92$  & $2.96$\footnotemark[1] & $1.094 \ (-15.6\%)$ & $1.355 \ (4.5\%)$ & $1.305 \ (0.6\%)$ & $1.323 \ (2.0\%)$ & $1.297$ & $ 1.09 \ (-15.5\%)$\footnotemark[4]\\  
    & VI  & $\delta$-MoC & $5.87$  & $1.86$\footnotemark[1] & $0.998 \ (-22.4\%)$ & $1.378 \ (7.2)\%$ & $1.256 \ (-2.2\%)$ & $1.376 \ (7.1\%)$ & $1.285$ & $ -$\\
\hline 
    & V   & TaC & $9.24$  & $3.06$\footnotemark[1] & $1.137\ (-12.1\%)$ & $1.319 \ (2.0\%)$ & $1.318 \ (1.9\%)$ & $1.321 \ (2.2\%)$ & $1.293$ & $1.14 \ (-12\%)$\footnotemark[5]\\
$5d$& VI  & WC  & $6.14$  & $-$    &$1.030 \ (-19.6\%)$ & $1.319 \ (4.1\%)$ & $1.254 \ (-1.1\%)$ & $1.354 \ (6.8\%)$ & $1.268$ & $ -$\\
\hline
\hline
\end{tabular}
\footnotetext[1]{The cleavage energies for the ($100$) surfaces are
  adapted from Ref.~\onlinecite{VinesTMC100} as
  $E_{\text{cleav}}=2E_{\text{surf}}$, since a stoichiometric ($100$)
  surface slab has two equivalent sides.}
\footnotetext[2]{Ref.~\onlinecite{AonoTiC}.}
\footnotetext[3]{Ref.~\onlinecite{RundgrenVC}.}
\footnotetext[4]{Ref.~\onlinecite{HayamiNbC}.}
\footnotetext[5]{Ref.~\onlinecite{HulbertTaC}.}
\end{center}
\end{table*}
%

The calculated relaxations of the four topmost layers of the
TMC($111$) surfaces are given in Table~\ref{tab:TMC_surf_relax}. A
pronounced contraction ($10$--$30\%$) of the first interlayer spacing
is found on all the surfaces. At the same time, the second interlayer
spacing increases compared to the bulk separation. Further down into
the slab the relaxations become smaller and the structure converges to
the bulk structure after three bilayers.  

Most of the TMC($111$) surfaces exhibit an alternating
positive-negative relaxation similar to the one found in
metals.\cite{JoMa88andHe94} Compared to the close-packed parent metal
surfaces, where the outer-layer relaxations are only a few percents of
the corresponding bulk interlayer
spacings,\cite{JoMa88andHe94,Feibelman} the relaxations of the
TMC($111$) surfaces are larger and comparable to other ionic polar
surfaces.\cite{RuYoLu02} This indicates that the bonding character in
TMC's has a significant ionic contribution, as discussed in
Ref.~\onlinecite{VoRu09}. The relaxation of the ($111$) surfaces is
hence quite different from the smaller rumpled relaxation found on the
non-polar ($100$) surfaces.   

The percentual relaxations of the ($111$) surfaces, relative to the
bulk interlayer spacings, decrease down each group. Along the periods,
the smallest percentual relaxation is found for group IV in period
$3d$ and group V in period $4d$. The larger percentual relaxations of
$\delta$-MoC and WC can be attributed to the NaCl structure being a
metastable phase for these compounds. The largest structural changes
are found on the ScC surface. The variations in percentual relaxations
are directly correlated to the cohesive energies:\cite{VoRu09} a small 
cohesive energy gives a large relaxation.         

For the contraction of the first interlayer distance there are some
experimental data, which agree qualitatively with our calculated
values and in most cases even quantitatively, as shown in
Table~\ref{tab:TMC_surf_relax}. For TiC($111$),
tight-binding~\cite{TanTiC111} and DFT~\cite{Kitchin} calculations
agree well with our values. Several of our results are also in
qualitative agreement with the theoretical ones in
Ref.~\onlinecite{ZhLiDiChZh02}. A deviation is found, however, for the
NbC($111$) surface, where experimental studies show a contraction of
both the first and the second interlayer distance, by $15.5\%$ and
$4\%$, respectively.\cite{HayamiNbC} The first observation agrees with
our calculated value but the second one differs qualitatively from the
contraction seen in our and other first-principles
studies.\cite{ZhangNbC111, Kitchin}

\subsection{Bader charge transfer}\label{sec:Bader_surf}
%
\begin{table}
\caption{
  The ionicity, that is, the amount of charge in units of $|e|$
  relative to the neutral atoms, of the atoms in the first (TM) and
  the second (C) surface layers (positive values = donated electrons,
  negative values = gained electrons) obtained from a Bader
  analysis. The percentual changes in ionicity compared to the
  bulk\cite{VoRu09} are also given (positive values = more
  electrons).}           
\label{tab:TMCs_surface_bader}
\begin{center}
\begin{tabular}{ccccccc}
\hline
\hline
Period & Group & Surface &  \multicolumn{2}{c}{TM} & \multicolumn{2}{c}{C}\\
       &       &         & rel.~atom & rel.~bulk& rel.~atom &rel.~bulk \\
\hline
     & III  & ScC &$+1.27$ & $+18\%$ & $-1.95$ &$+27\%$ \\
$3d$ & IV   & TiC &$+1.08$ & $+28\%$ & $-1.77$ &$+19\%$ \\
     & V    & VC  &$+0.86$ & $+39\%$ & $-1.47$ &$ +4\%$ \\
\hline
     & IV   & ZrC &$+1.15$ & $+32\%$ & $-1.93$ &$+14\%$ \\
$4d$ & V    & NbC &$+0.86$ & $+48\%$ & $-1.58$ &$ -4\%$ \\
     & VI   & $\delta$-MoC & $+1.39$& $+29\%$& $-2.29$ &$+16\%$ \\
\hline
     & V    & TaC &$+1.07$ & $+45\%$ & $-1.93$ & $-1\%$  \\
$5d$ & VI   & WC  &$+0.86$ & $+46\%$ & $-1.53$ & $-4\%$  \\
\hline
\hline
\end{tabular}
\end{center}
\end{table}
%

In the bulk TMC systems there is a charge transfer from the TM to the
C atoms.\cite{VoRu09} Table~\ref{tab:TMCs_surface_bader} shows our
calculated Bader charge values for the TM-terminated TMC($111$)
surfaces. Both the values relative to the free atoms (in units of
electronic charge $|e|$) and the values relative to the bulk values
(in percentages) are given.  

Compared to the bulk, an accumulation of charge occurs on the first
surface bilayer, mainly on the first TM atomic layer but also in
several cases on the topmost C atomic layer. An exception is
ScC($111$), where the C layer gains more electrons than the Sc
layer. For all the investigated TMC's there is a total of $40$--$50\%$
reduction in the surface bilayer ionicity compared to the
corresponding bulk ionicity.\cite{VoRu09} For the NbC($111$) surface
our ionicity value for the surface Nb atoms ($+0.86 e$) is in good
agreement with the calculated one of $+0.90 e$ by Zhang \textit{et
  al.}\cite{ZhangNbC111}                 

To the right along each period (except for $\delta$-MoC), the charge
accumulation on the first TM layer atoms increases, while it decreases
on the C layer atoms. The same trends are found down each group. 

The calculated charge accumulation at the TMC($111$) surfaces has its
origin in the polar nature of these surface. As mentioned above, the
TMC's are partially ionic materials. Therefore there is a macroscopic
electric field caused by the non-zero perpendicular dipole moment of
the TMC bilayer, which makes the ($111$) surface polar and
unstable.\cite{Ta79} To counteract this, a surface charge can be
induced, creating a neutralizing electric field. In
Ref.~\onlinecite{TsuHo82}, it is shown that for the ($111$) surface of
a crystal with the NaCl structure, this surface charge should be equal
to $50\%$ of the bulk ionicity, which is what we observe. 

The extra charge in the surface bilayer will affect its electronic
structure and be of importance for its adsorption characteristics, as
described below in Sections~\ref{sec:surf_el_struc} and \ref{sec:ads},
respectively.

\subsection{Electronic structure}\label{sec:surf_el_struc}

In this Section, we investigate in detail the electronic structure of
the clean TMC(111) surfaces to gain an understanding of which
characteristics that are surface unique and potentially important for
the adsorption.

\subsubsection{Bulk characteristics}

For convenience and to facilitate the discussion of the TMC($111$)
surfaces, we here provide a very short summary, based on
Ref.~\onlinecite{VoRu09}, of some bulk characteristics of the TMC's. 

The bonding in bulk TMC's has contributions from iono-covalent TM--C
bonds, from TM--TM bonds, and from C--C bonds. For all considered
TMC's, the bulk DOS's and band structures consist of a low-lying
valence band (LVB), dominated by C($2s$) states, an upper valence band
(UVB), with contributions from C($2p$) and TM($d$) states, and a
conduction band (CB) of mainly TM($d$) character. The covalent
TM($d$)--C($2p$) bonding states are found in the UVB. The main
contribution to the UVB (CB) comes from C($2p$) [TM($d$)] states, thus
indicating the partially ionic character of the bond. The UVB and the
CB, positioned on each side of $E_F$, are connected by a non vanishing
continuum of TM--TM and TM--C states (for ScC also C--C states)
states. The C($2p$)--C($2p$) bonds are found in the lower part of the
UVB.   

Towards the right along each period, both the UVB and the CB are
shifted down in energy relative to $E_F$ and the UVB becomes less C
localized. Down each group, the bands are shifted towards lower
energies and the UVB becomes more C localized.

\subsubsection{Common surface characteristics}

\begin{figure*}
  \centering
  \includegraphics[width=\textwidth]{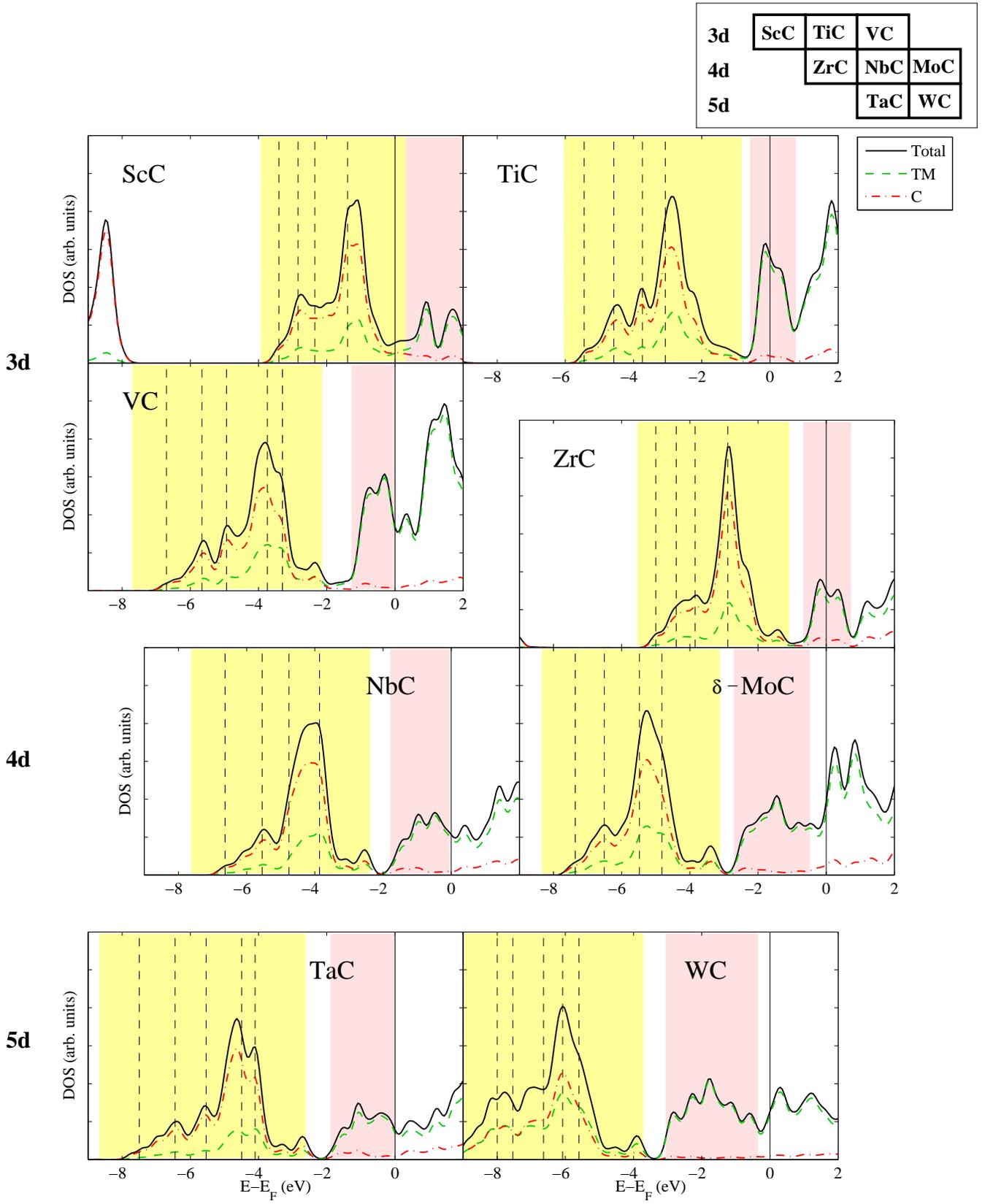}
  \caption{\label{fig:DOS_surface}
    Total and atom projected densities of states for the first surface
    bilayer of the considered TMC($111$) surfaces. The shaded yellow
    areas indicate the bulk UVB regions and the vertical dashed lines
    mark the positions of the CSR's. The TMSR's are marked with shaded 
    red areas.}      
\end{figure*}

\begin{figure}
  \centering
  \includegraphics[width=0.5\textwidth]{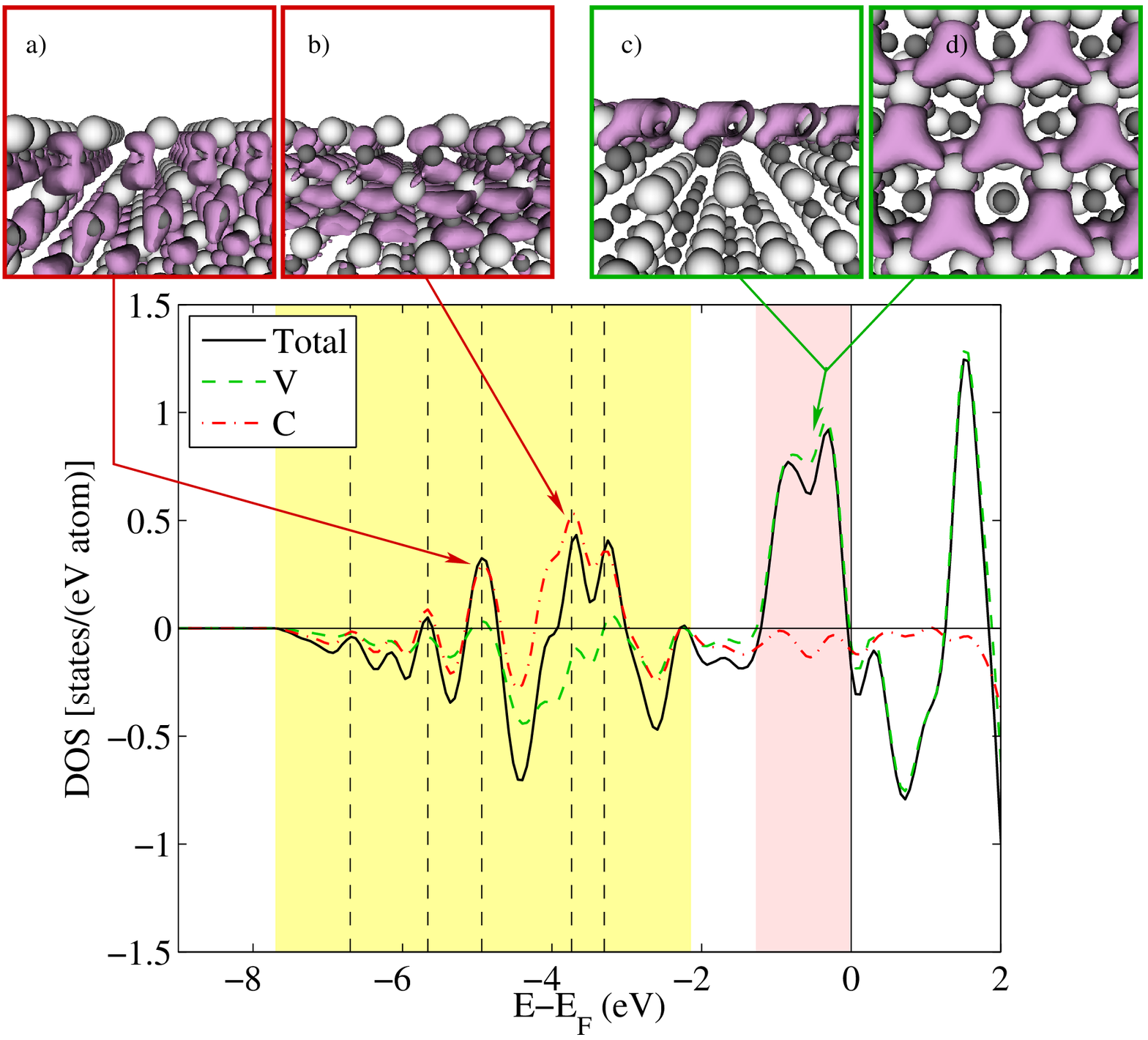}
  \caption{\label{fig:diff_surf_bulk}
    Difference between the surface and bulk DOS's for the first
    surface bilayer of VC($111$). The shaded areas and vertical dashed
    lines represent the same quantities as in
    Fig.~\ref{fig:DOS_surface}. Also shown are representative
    real-space Kohn-Sham wave functions illustrating the existence of
    (a-b) CSR's localized on the first bilayer C atoms at $-4.9$~V and
    $-3.7$~eV, respectively and (c-d) a TMSR localized on the first
    bilayer V atoms. Figures (a-c) are side views and (d) is a top
    view of the surface. Larger gray balls correspond to V atoms and
    smaller black balls are C atoms.}              
\end{figure}

The total and atom projected DOS's of the first TMC surface bilayer
are presented in Fig.~\ref{fig:DOS_surface}. A common feature of the
bulk and surface DOS's is that both have an LVB, a UVB, and a CB. The
bulk DOS is recovered in the third surface bilayer. Despite these
similarities, all surface DOS's differ sharply from their respective
bulk DOS's. To easier identify the surface properties we study the
difference between the surface DOS (Fig.~\ref{fig:DOS_surface}) and
the bulk DOS (adapted from our bulk study~\cite{VoRu09}), as
illustrated for VC in Fig.~\ref{fig:diff_surf_bulk}. 

For all considered TMC ($111$) surfaces, surface specific features 
similar to those found on TiX($111$)\cite{RuLu07,RuVoLu06,RuVoLu07}
can be identified, in particular:  
(i) a TM-localized surface resonance (TMSR), positioned in the bulk
pseudogap at or in the vicinity of $E_F$ (shaded red area in
Figs.~\ref{fig:DOS_surface} and \ref{fig:diff_surf_bulk}) and  
(ii) a more strongly C-localized UVB than in the bulk, with several
C-localized surface resonances (CSR's) in the lower part of the UVB
(shaded yellow area in Figs.~\ref{fig:DOS_surface} and
\ref{fig:diff_surf_bulk}). The energetical extent of the surface UVB
is, however, largely similar to that of the bulk UVB.  

Both kinds of surface resonances consist of states localized at the
surface that overlap energetically with bulk states, hence the term
surface resonance. They can be identified as positive peaks in the
DOS-difference plots and from real-space analyses of the Kohn-Sham
wave functions (Fig.~\ref{fig:diff_surf_bulk}). These show that the
TMSR's correspond to unsaturated TM bonds that extend into the vacuum
[see Fig.~\ref{fig:diff_surf_bulk}(c)] and point towards the fcc sites
[see Fig.~\ref{fig:diff_surf_bulk}(d)] where the C atoms would be
present if the bulk stacking in the ($111$) direction had been
continued. Hence the TMSR are dangling bonds with a three-fold
symmetry. The C-localized states in the surface bilayer rearrange to
form new bonds; however, they do not become fully saturated, which
results in C dangling bonds striving in between the first-layer TM
atoms towards the vacuum [see
  Fig.~\ref{fig:diff_surf_bulk}(a-b)]. Coupling of these states to the
second bilayer bulk states gives them the status of surface
resonances.             

In addition, several negative peaks can be identified in the
DOS-difference plots (Fig.~\ref{fig:diff_surf_bulk}):  
(i) one region of mixed TM and C character, positioned in the upper 
part of the UVB and just below the TMSR (between the shaded yellow and
red regions in Fig.~\ref{fig:diff_surf_bulk}); 
(ii) several peaks of mixed TM and C character, positioned in between
the CSR peaks in the lower part of the UVB;
and  
(iii) one region of mainly TM character, positioned in the lower part
of the CB. These regions of negative peaks correspond to strong
quenchings of the bulk UVB and CB peaks, respectively.   

%
%

%
%

\subsubsection{Trends in surface characteristics} 

The analysis of the difference between surface and bulk DOS's shows
that as the group number of the surface TM constituent increases along
each period, the position of the TMSR is shifted to lower energies
relative to $E_F$, as expected from the filling of the TM $d$
states. The TMSR filling increases, varying from a non-filled TMSR
above $E_F$ for group III, to a partly filled TMSR at $E_F$ for group
IV, to a filled TMSR just below $E_F$ for group V, and a filled low
lying TMSR for group VI. At the same time, the TMSR amplitude
decreases, while the width is constant or increases slightly. 

A shift to lower energies and a decreasing amplitude are observed also
for the CSR's, starting from ScC($111$), whose CSR's lie just below
$E_F$. In addition, the amplitudes of the negative DOS-difference
peaks in the upper part of the UVB and in the lower part of the CB
decrease. For ScC, both the DOS-difference plots and the Kohn-Sham
wave functions show more pronounced CSR's than on the other TMC's. 

The DOS-difference plots show that as the period number of the surface
TM constituent increases down each group, the positions of the TMSR
and of the CSR's are more or less unaffected, while the TMSR amplitude
decreases and the width increases slightly.

\subsubsection{Connection with experiments}

Several of the TMC($111$) surfaces have been studied by means of angle
resolved photoemission spectroscopy (ARPES), revealing
surface-localized states, there denoted ``surface
states''.\cite{WeaverTiC,BrandshawTiC,ZaimaTiC,EdamotoTiC} We
associate these with our TMSR's, termed ``resonances'' because of
their location in the pseudogap where the bulk DOS is non
vanishing. The ARPES study, with an experimental resolution of
$0.2$~eV, places the position of these TMSR's in the $\Gamma$-point at
$-0.2$~eV for TiC,~\cite{WeaverTiC,BrandshawTiC,ZaimaTiC,EdamotoTiC}
at $-0.2$~eV for ZrC,~\cite{EdamotoZrC} at $-0.7$~eV for
NbC,\cite{EdamotoNbC} and at $-0.7$~eV for TaC,\cite{AnazawaTaC} in
perfect agreement with our results.

\subsection{Origin of the SR's}

The appearance of the TMSR's is a result of the breakage of the
iono-covalent TM--C bonds that cross the ($111$) cleavage plane, which
causes (i) anti-bonding TM states in the empty bulk CB to collapse
into more TM atomic-like states, which lie at a lower energy than the
CB and are positioned at or in the vicinity of $E_F$, and (ii) bonding
C states in the upper part of the UVB to vanish as C atoms are removed
from the surface. This is seen in our DOS-difference plots
(illustrated by Fig.~\ref{fig:diff_surf_bulk}) as negative peaks in
the upper part of the bulk UVB region and in the lower part of the
bulk CB region. A similar finding was reported for
TiC($111$).\cite{RuLu07,RuVoLu07} The existence of TMSR's supports the
charge accumulation picture obtained from the Bader analysis.   

In an analogous way, the changes in the lower part of the UVB DOS can
be interpreted to be due to the breakage of the bulk C--C bonds upon
formation of the surface, which results in the formation of CSR's. The
increase of Bader charge on the C atoms in the surface bilayer
supports the existence of CSR's and arises partly due to the influence
of the extra charge on the TM surface bilayer on the C atoms.  

The increase in amount of extra charge on the TM surface layer,
compared to the bulk, that takes place when moving to the right along
a period (see Section~\ref{sec:Bader_surf}) is due to the filling of
the TM $d$ states and therefore of the TMSR. At the same time the
amount of extra charge on the C layer, compared to the bulk,
decreases.  

For the ScC surface the whole DOS is positioned at higher energies
compared to the other TMC's. We recall that in bulk ScC, the C--C
bonds are more pronounced than in the other TMC's.\cite{VoRu09} In
addition, the energy separation between UVB and CB is largest for this
TMC and the Sc-localized CB is very high up in energy relative to
$E_F$. Therefore the ScSR states are not populated. However, the
surface still strives to reduce its polarity, which is done by a
charge accumulation on the C atoms, resulting in very pronounced
CSR's.


\section{Trends in atomic adsorption}\label{sec:ads}

So far we have discussed the properties associated with clean TMC
surfaces. In this Section the focus is on the atomic adsorption on the
TM-terminated TMC($111$) surfaces. Atomic adsorption is one of the
first necessary processes in reactions at surfaces. We perform two
series of trend studies: one with respect to the adsorbate and one
with respect to the substrate. As adsorbates we consider period 1 and
period 2 atoms H, B, C, N, O, and F. The substrate trend involves the
change of the TM atom in the TMC. In our case we use the carbides ScC,
TiC, VC, ZrC, NbC, $\delta$-MoC, TaC and WC, thus spanning four groups
and three periods in the periodic table (see
Fig.~\ref{fig:def_TMC}). The consequences of changing the non-metal
atom have been investigated in previous studies, where adsorption on
TiC($111$) and TiN($111$) was
conducted.\cite{VoRuLu06,RuVoLu06,RuVoLu07,RuLu07}

\subsection{Computational details}

The systems are modeled by four bilayers of TMC with a $3$ by $3$ atom
geometry in the surface plane. We use a cutoff energy of $400$~eV and
a Monkhorst-Pack sampling of $4\times 4\times1$ $k$ points. The high
symmetry adsorption sites fcc, hcp, top, and bridge are
considered. Both the adatom and the three top-most surface bilayers
are allowed to relax in all directions.       

The adsorption energies $E_{\text{ads}}$ are defined as 
\begin{equation}
E_{\text{ads}} = -(E_{\text{slab+adatom}} - E_{\text{clean slab}} - E_{\text{free adatom}}),
\end{equation}
where $E_{\text{free adatom}}$ is the energy of an isolated
spin-polarized atom. It is known that the $E_{\text{ads}}$ values are
sensitive to the choice of GGA flavor. To get an understanding of the
flavor dependence we have performed calculations on TMC($111$) systems
with adsorbed H, N, and O atoms using the RPBE GGA
functional.\cite{RPBEHammer} The RPBE functional has been shown to
give better results than the PW91 GGA functional for adsorption
energies on TM surfaces,~\cite{RPBEHammer} whereas for bulk structure
determination in some cases it gives worse results than the PW91
functional.\cite{KuPeBla99} According to our calculations on H, N, and
O adsorbed on TMC($111$) surfaces, the RPBE functional gives
consistently lower $E_{\text{ads}}$ values than the PW91 functional. A
similar type of variation was found for adsorption on pure TM
surfaces.\cite{RPBEHammer} As the decrease in $E_{\text{ads}}$ values
for each adsorbate is the same on all the considered TMC($111$)
surfaces ($0.32$~eV, $0.45$~eV, and $0.12$~eV for the H, N, and O
adsorbates, respectively), the $E_{\text{ads}}$ trends with respect to
the substrate are not affected by the choice of GGA functional.  Also,
as these changes are smaller than $0.5$~eV, the changes in
$E_{\text{ads}}$ trends with respect to the adsorbate are not
significant enough to affect the key features of these trends.

\subsection{Atomic adsorption energies}\label{sec:Eads}

%
\begin{table}
\caption{\label{tab:ads_TMXs}
Calculated atomic adsorption energies in eV. The preferred adsorption
site is fcc, except for C, N, and O on $\delta$-MoC and N and O on WC,
which favor hcp site.}  
\begin{center}
\begin{tabular}{ccccccccc}
\hline
\hline
Period & Group & Surface (site)&  H    &  B    &  C    &  N    &  O    &  F   \\
\hline
       & III   & ScC (fcc)    & $2.97$& $2.92$& $4.52$& $5.36$& $7.68$& $6.97$ \\
$3d$   & IV    & TiC (fcc)    & $3.60$& $5.68$& $7.87$& $7.86$& $8.75$& $6.92$ \\
       & V     & VC  (fcc)    & $3.29$& $5.83$& $7.54$& $6.85$& $7.31$& $5.79$ \\
\hline
       & IV    & ZrC (fcc)    & $3.62$& $5.56$& $7.73$& $7.79$& $8.73$& $7.00$ \\
$4d$   & V     & NbC (fcc)    & $3.45$& $5.89$& $7.61$& $6.92$& $7.41$& $5.85$ \\
       & VI    & $\delta$-MoC (fcc) & $3.08$& $5.70$& $6.76$& $5.61$& $6.33$& $5.10$ \\
       &       & $\delta$-MoC (hcp) & $2.81$& $5.60$& $7.11$& $6.19$& $6.54$& $5.09$ \\
\hline
       & V     & TaC (fcc)    & $3.48$& $5.95$& $7.73$& $7.10$& $7.53$& $5.71$ \\ 
$5d$   & VI    & WC (fcc)     & $2.95$& $5.84$& $7.23$& $6.12$& $6.40$& $4.95$ \\
       &       & WC (hcp)     & $2.69$& $5.73$& $7.22$& $6.32$& $6.59$& $4.89$ \\
\hline
\hline
\end{tabular}
\end{center}
\end{table}
%

%
\begin{figure*}
\centering  
\includegraphics[width=0.49\textwidth]{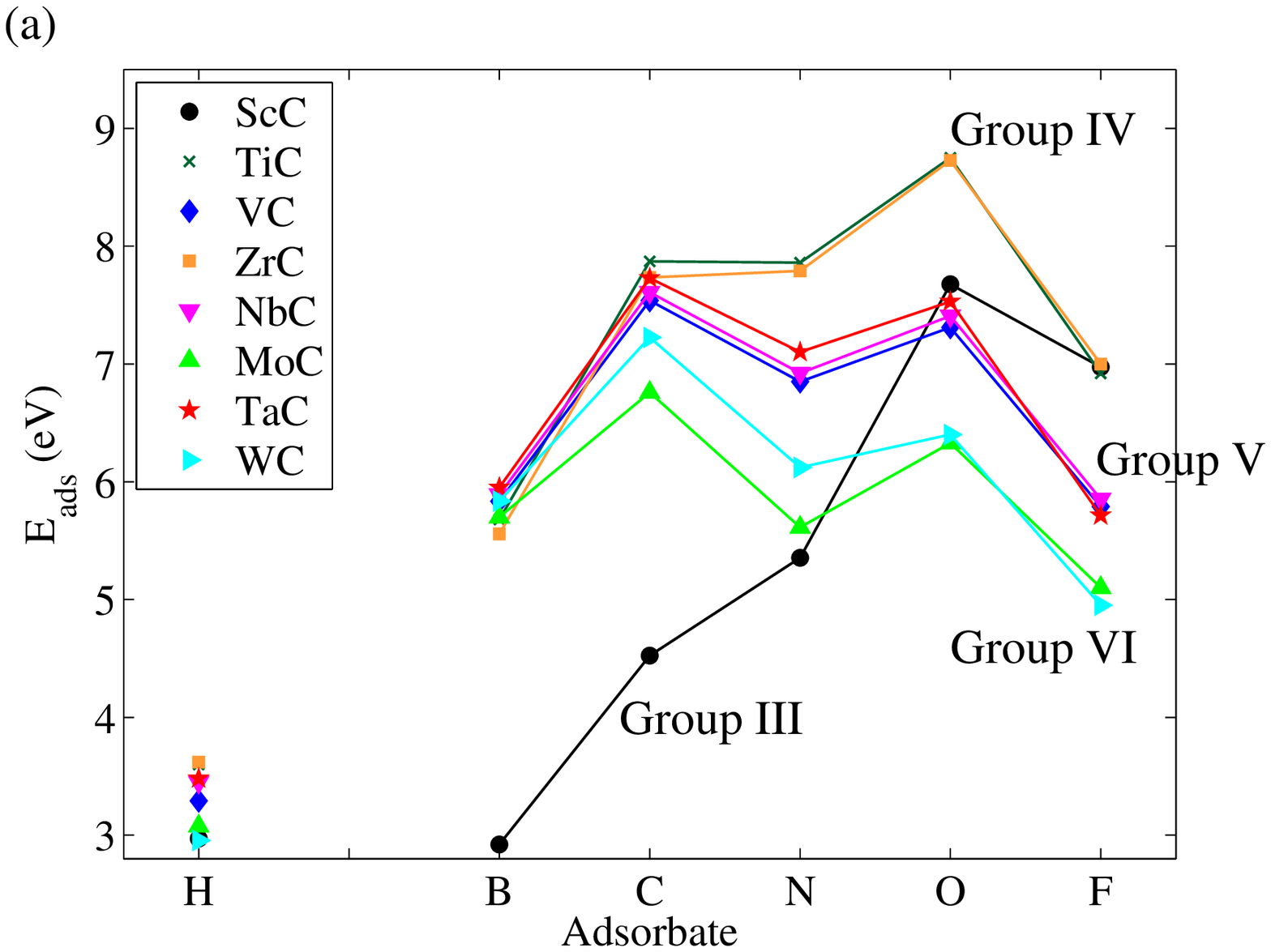}
\includegraphics[width=0.49\textwidth]{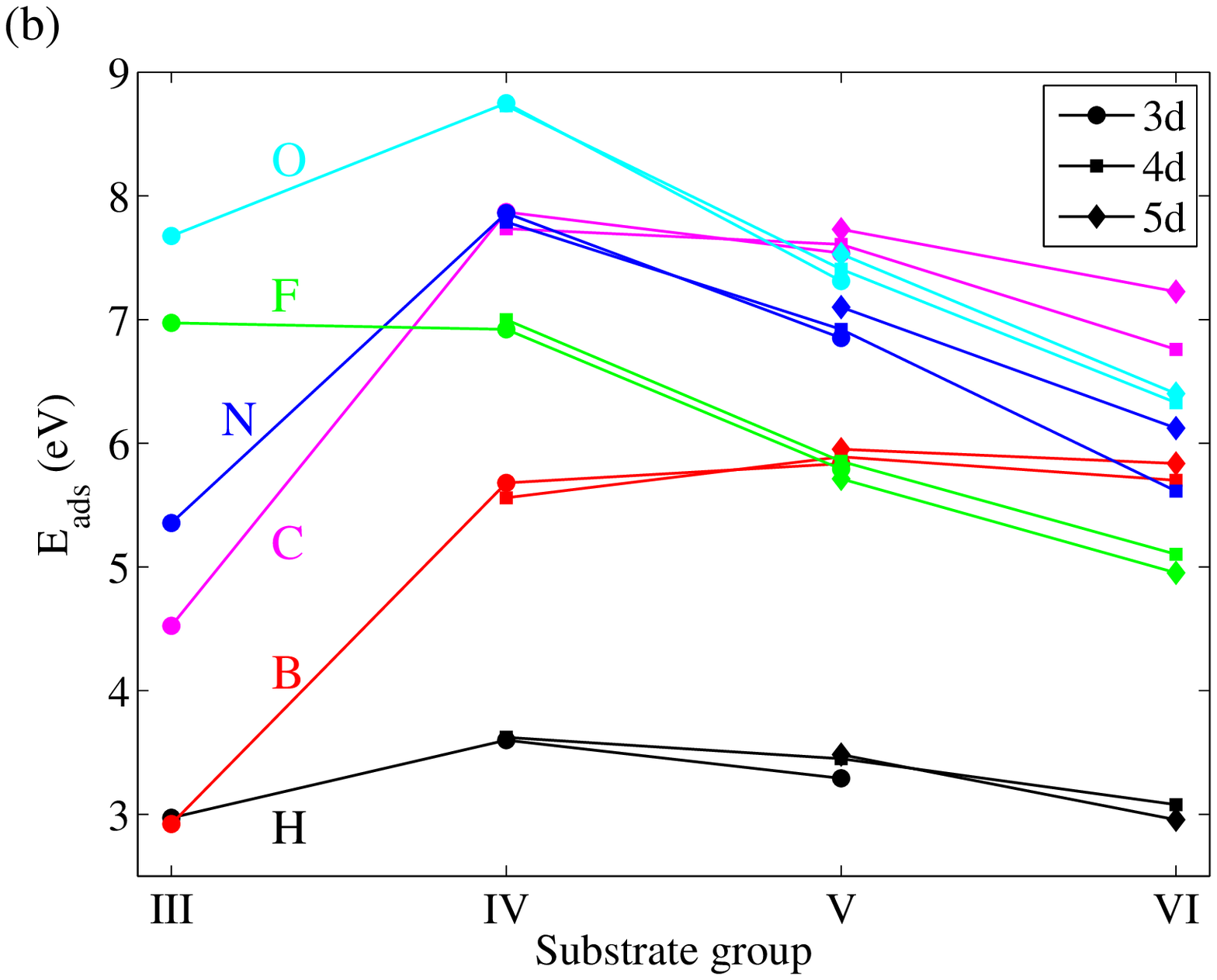}
\caption{\label{fig:Ads_on_TMX_all_ads_sub}
  Calculated adsorption energies $E_{\text{ads}}$ for period 1 and
  period 2 adatoms in fcc site on the considered TMC($111$) surfaces
  as a function of (a) the adsorbate atom and (b) the group number of
  the substrate TM constituent.}      
\end{figure*}
%

To elucidate our results in a clear way we gather the calculated
$E_{\text{ads}}$ values in Table~\ref{tab:ads_TMXs} and
Fig.~\ref{fig:Ads_on_TMX_all_ads_sub}. The most favorable adsorption 
site is fcc, except for some adsorbates on $\delta$-MoC and WC, which
favor the hcp site, closely followed by the fcc site. The relaxed
perpendicular distances between each adatom and the TMC(111)
substrates are given in Table~\ref{tab:ads_geom}. In the following,
the discussion will focus on the fcc site. There are three arguments 
for this choice. First of all the shape of the $E_{\text{ads}}$ trends
does not change if the hcp preference is considered. Secondly, as we
are interested in focusing on the changes in surface electronic
structure due to electronic factors we want to reduce the effects
arising from geometric factors. Thirdly, in Ref.~\onlinecite{RuLu07}
it has been shown for TiC($111$) that a similar adsorption mechanism
applies for adsorption in both fcc and hcp sites.  

When moving from left to right along the adatom period 2
[B$\rightarrow$F, see Fig.~\ref{fig:Ads_on_TMX_all_ads_sub}(a)] we
find  
(i) ``M''-shaped $E_{\text{ads}}$ trends for all TMC's;  
(ii) grouping, that is, similar $E_{\text{ads}}$ values within each 
substrate group; and
(iii) strongest adsorption either for O or for C, depending on the 
substrate. The smallest variations in $E_{\text{ads}}$ values are
found on $\delta$-MoC.  

For each adsorbate, the trends in $E_{\text{ads}}$ along the substrate
periods [Fig.~\ref{fig:Ads_on_TMX_all_ads_sub}(b)] show that 
as group IV $\rightarrow$ VI, 
(i) the adsorption strength decreases for the adsorbates N, O, and F;   
(ii) the $E_{\text{ads}}$ variations are small for H and C; and 
(iii) the adsorption strength increases slightly for B. The strongest
adsorption is found on group IV TMC's (except for B).  

On ScC, the adsorption trends show somewhat different behaviors than
on the other substrates. Also, the variations in $E_{\text{ads}}$
within each substrate group are much smaller than the ones within each
substrate period.

\begin{table}
\centering
\caption{\label{tab:ads_geom}
Calculated perpendicular distances in \AA\ between the adatoms and the
TMC($111$) substrates for the systems given in
Table~\ref{tab:ads_TMXs}.}     
\begin{tabular}{ccccccccc}
\hline\hline
Period & Group & Surface (site)    &  H    &  B    &  C    &  N    &  O    &  F   \\
\hline
       & III   & ScC (fcc)         & $1.25$& $1.55$& $1.25$& $1.07$& $1.09$& $1.30$ \\
$3d$   & IV    & TiC (fcc)         & $1.03$& $1.32$& $1.10$& $1.02$& $1.07$& $1.28$ \\
       & V     & VC  (fcc)         & $0.99$& $0.97$& $1.11$& $1.07$& $1.14$& $1.38$ \\
\hline
       & IV    & ZrC (fcc)         & $1.01$& $1.35$& $1.11$& $1.03$& $1.08$& $1.31$ \\
$4d$   & V     & NbC (fcc)         & $1.03$& $1.33$& $1.15$& $1.10$& $1.17$& $1.42$ \\
       & VI    & $\delta$-MoC (fcc) & $1.16$& $1.36$& $1.25$& $1.30$& $1.46$& $1.62$ \\
       &       & $\delta$-MoC (hcp) & $1.24$& $1.35$& $1.26$& $1.24$& $1.34$& $1.59$ \\
\hline
       & V     & TaC (fcc)         & $1.00$& $1.32$& $1.14$& $1.08$& $1.14$& $1.36$ \\ 
$5d$   & VI    & WC (fcc)     & $1.06$& $1.33$& $1.17$& $1.14$& $1.21$& $1.50$ \\
       &       & WC (hcp)     & $1.21$& $1.34$& $1.19$& $1.17$& $1.25$& $1.56$ \\

\hline
\hline
\end{tabular}
\end{table}

\subsection{Bader analysis} 

The Bader analysis gives a fractional electronic charge transfer from
the substrate to the adsorbate (see
Table~\ref{tab:Bader_TMXs_ads}). This indicates that the
adsorbate--TMC bond is partially ionic. Most of the charge that the
adsorbate gains is from the nearest three surface TM atoms, except on
ScC where the C atoms are the main electron donors.       

The variations in the charge transfer down the TMC groups are much
smaller than those along the TMC periods. This indicates a more
similar adsorbate--substrate interaction within substrate groups than
within substrate periods. The largest amounts of charge transfer are
found for the adsorbates N and C. However, the most filled outer
electron shell is found for F, which is the most electronegative of
the adsorbates. In our previous work on TiX($111$) we found, based on
Bader analysis and the adsorbate-induced changes in electron density,
that F has the strongest ionic adsorbate--substrate
bond.\cite{RuLu07,VoRuLu06,RuVoLu07} 

Overall, there is no clear correlation between the calculated
$E_{\text{ads}}$ and charge transfer values, implying that the charge
transfer alone cannot explain the $E_{\text{ads}}$ trends. In the same
way as in our previous studies on
TiX($111$),\cite{RuLu07,VoRuLu06,RuVoLu07} we show in the following
that there is a significant covalent contribution to the
adsorbate--TMC($111$) bond.  

%
\begin{table}
\centering
\caption{
  Calculated Bader charge transfers, in units of electron charge
  $|e|$, from the TMC($111$) surface to the atomic adsorbate for the
  systems given in Table~\ref{tab:ads_TMXs}.}   
\label{tab:Bader_TMXs_ads}
\begin{tabular}{ccccccccc}
\hline
\hline
Period & Group & Surface (site) &  H  &  B  &  C  &  N  &  O  &  F  \\
\hline
     & III & ScC (fcc) & $0.74$ & $1.05$ & $1.45$ & $1.58$ & $1.33$ & $0.84$ \\
$3d$ & IV  & TiC (fcc) & $0.64$ & $1.09$ & $1.35$ & $1.35$ & $1.17$ & $0.80$ \\
     & V   & VC  (fcc) & $0.58$ & $0.76$ & $1.16$ & $1.21$ & $1.09$ & $0.76$ \\
\hline
     & IV & ZrC (fcc) & $0.67$ & $1.19$ & $1.49$ & $1.50$ & $1.25$ & $0.82$ \\
$4d$ & V  & NbC (fcc) & $0.59$ & $0.92$ & $1.29$ & $1.33$ & $1.18$ & $0.80$ \\
     & VI & $\delta$-MoC (fcc)  & $0.65$ & $1.16$ & $1.55$ & $1.54$ & $1.26$ & $0.82$ \\
     &    & $\delta$-MoC (hcp)  & $0.58$ & $1.10$ & $1.51$ & $1.55$ & $1.27$ & $0.82$ \\
\hline
     & V  & TaC (fcc) & $0.63$ & $1.04$ & $1.44$ & $1.45$ & $1.25$ & $0.81$ \\ 
$5d$ & VI & WC (fcc)  & $0.70$ & $1.58$ & $2.14$ & $1.96$ & $1.50$ & $0.88$ \\
     &    & WC (hcp) & $0.61$ & $1.42$ & $2.10$ & $1.97$ & $1.48$ & $0.85$ \\
\hline
\hline
\end{tabular}
\end{table}
%

\subsection{Density of states}\label{sec:DeltaDOS}

To learn about, in particular, the covalent parts of the adsorption
bond, this Section is devoted to the trends in the adsorption-induced
electronic structure. More precisely, the difference in the surface
DOS before and after adsorption ($\Delta$DOS), that is, the
adsorbate-induced changes in DOS are investigated. With this very
useful tool we monitor trends with respect to both substrate and
adsorbate.

\subsubsection{Common characteristics in $\Delta$DOS for atomic adsorbates} 

Figures~\ref{fig:C_DeltaDOS_all_substrates} and
\ref{fig:O_DeltaDOS_all_substrates} show the $\Delta$DOS's for C and
O, respectively, adsorbed on the considered TMC($111$)
surfaces. Figure~\ref{fig:VC_DeltaDOS_all_adsorbates} shows the
$\Delta$DOS's for the different adsorbates on the VC($111$)
surface. The general form of all the $\Delta$DOS's consists of
negative peaks of exclusively TM $d$ character at the location of the
clean-surface TMSR's (the shaded red regions) and negative peaks (or
minima) of C character at the location of the clean-surface CSR's
(vertical lines in the shaded yellow regions). Positive $\Delta$DOS
peaks are observed below, in between, and above the various negative
peaks. A more detailed analysis of the atom-projected $\Delta$DOS's
shows that all the TMSR states associated with the three TM atoms
close to the fcc site, where the adatom is adsorbed, are depleted.

\subsubsection{Trends in $\Delta$DOS with respect to the substrate}\label{sec:substrateDOSs} 

%
\begin{figure*}
\centering
\includegraphics[width=\textwidth]{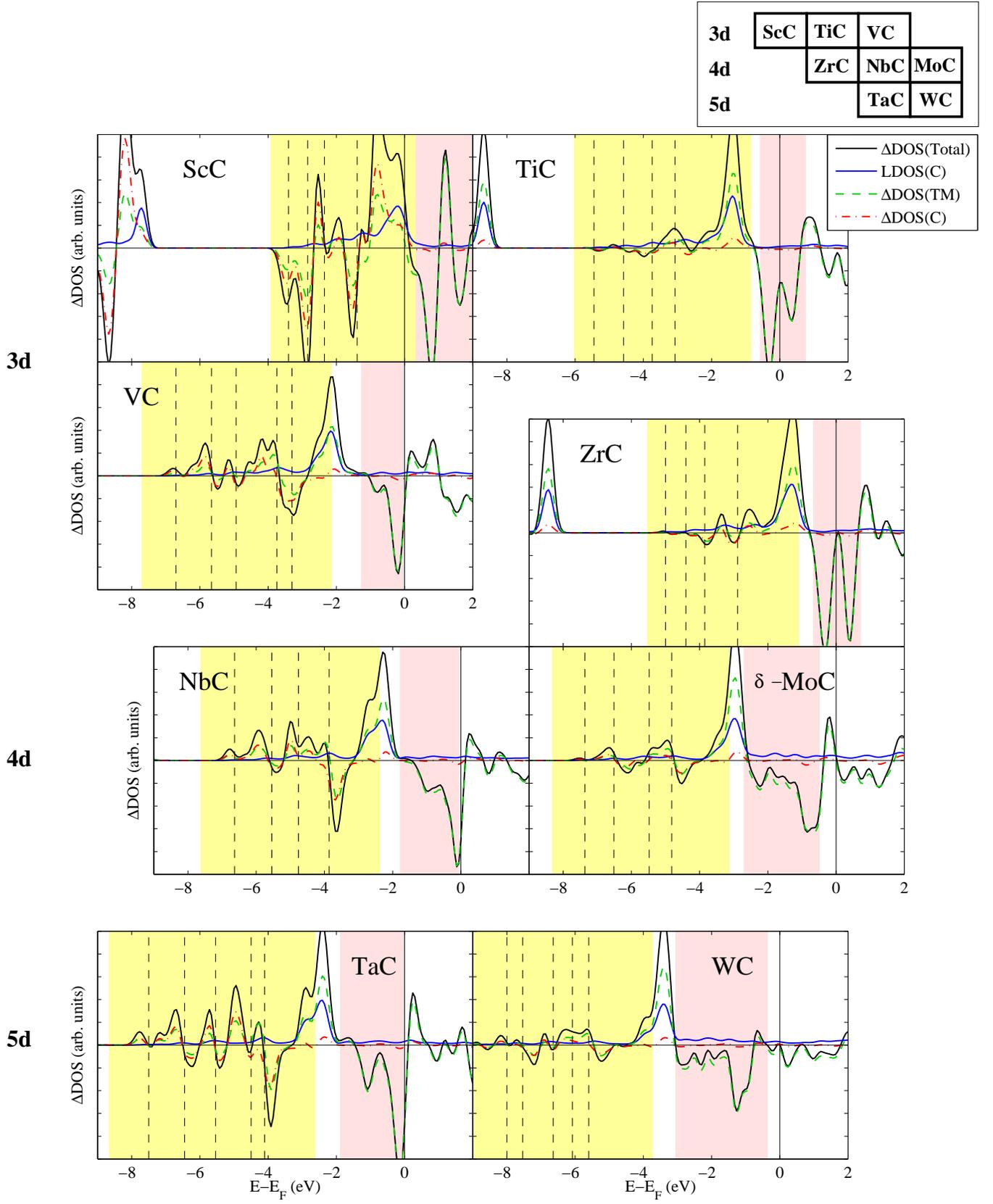}
\caption{\label{fig:C_DeltaDOS_all_substrates}
  Total and atom-projected $\Delta$DOS's for the adsorbate C atom in
  fcc site on the considered TMC($111$)surfaces. The shaded areas and
  the dashed vertical lines represent the same quantities as in
  Fig.~\ref{fig:DOS_surface}.}         
\end{figure*}
%

%
\begin{figure*}
\centering
\includegraphics[width=\textwidth]{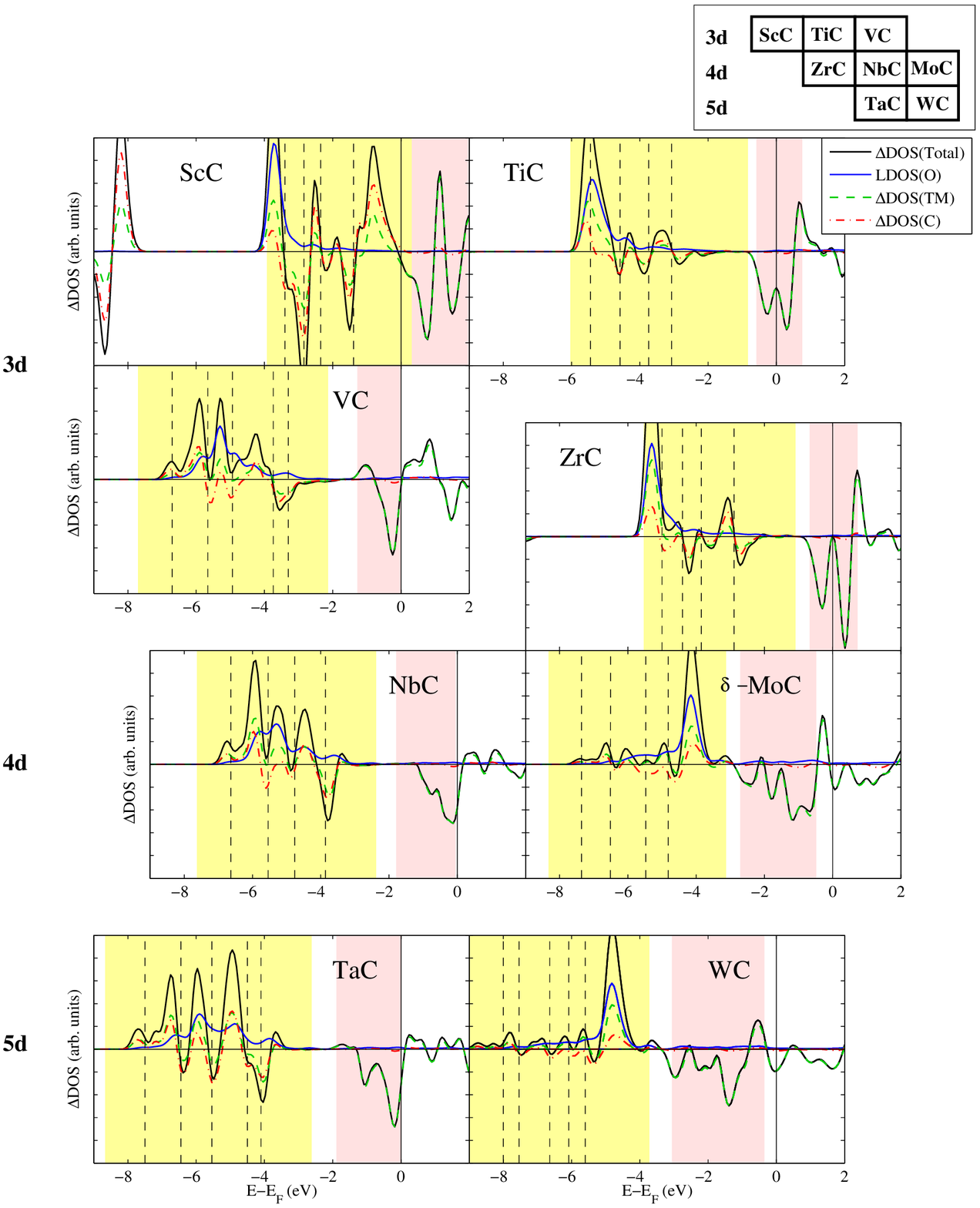}
\caption{\label{fig:O_DeltaDOS_all_substrates}
  Total and atom-projected $\Delta$DOS's for the adsorbate O atom in
  fcc site on the considered TMC($111$) surfaces. The shaded areas and
  the dashed vertical lines represent the same quantities as in
  Fig.~\ref{fig:DOS_surface}.}       
\end{figure*}
%

The variations in $\Delta$DOS between the different substrates are
illustrated by addressing the differences between two representative
example: C and O atoms, respectively.

The $\Delta$DOS for adsorbed C,
Fig.~\ref{fig:C_DeltaDOS_all_substrates}, shows the presence on all
substrates of a pronounced positive adsorbate-projected DOS
(represented by the blue line) peak that is pinned just below the
depleted TMSR region and that extends with a low-amplitude tail
throughout the upper part of the UVB. This peak overlaps with
substrate-TM states and can therefore be identified as a bonding
adsorbate--TMSR level. The corresponding antibonding level, of mainly
TM character, is found above the depleted TMSR region. The $\Delta$DOS
in the UVB consists of a number of subpeaks, of mixed adatom-C,
substrate-TM, and substrate-C character, that are located in between
the depleted CSR levels. Such a structure indicates that the bonding
adsorbate--TMSR level has interacted with the CSR's in the UVB.  

The ScC substrate forms an exception to the above pattern. Here, the
interaction with the CSR's in the UVB appears to be much stronger.
Also, the overlap of the adsorbate-projected DOS peak with the
substrate-TM states is not as pronounced as on the other TMC
substrates.  

The $\Delta$DOS for adsorbed O,
Fig.~\ref{fig:O_DeltaDOS_all_substrates}, shows a much more varied
structure. On several substrates, the adsorbate-projected DOS is now
located inside the UVB region and shows a stronger substrate-C
character than in the case of adsorbed C. When moving from left to
right along each TMC period, this peak shifts to higher energies
relative to the substrate UVB region: for group IV TMC's (TiC and ZrC)
it lies at the lower edge of the UVB, whereas for group VI TMC's
($\delta$-MoC and WC) it lies at the upper edge of the UVB. At the
same time, its form changes: quite localized with a tail towards
higher energies for TiC and ZrC; broad for VC, NbC, and TaC; and
localized with a tail towards lower energies for $\delta$-MoC and
WC. These observations indicate a stronger and much more varied
interaction between the bonding adsorbate--TMSR level and the
substrate CSR's than in the case of adsorbed C. 

Again, ScC stands out from the above pattern, being characterized by a
much stronger participation of the substrate CSR's in the bonding.  

It can also be noted that for both adsorbed C and adsorbed O, the key
features of the $\Delta$DOS's are the same within each substrate TM
group. 

These results show the presence of significant covalent bondings
between the adsorbate and the substrate SR's.  In particular, both
types of SR's (TMSR's and CSR's) appear to participate in the
chemisorption.  This indicates that the previously reported picture
for adsorption on TiC($111$) and TiN($111$), that is, a
concerted-coupling model (CCM) should be valid on these TMC($111$)
surfaces as well. Such a picture is further pursued in  
Section~\ref{sec:discussion}, where the differences and variations in
electronic structure are related to the trends in calculated
$E_{\text{ads}}$ values described in Section~\ref{sec:Eads}.

\subsubsection{Trends in $\Delta$DOS with respect to the adsorbate} 

%
\begin{figure}
\centering
\includegraphics[width=0.5\textwidth]{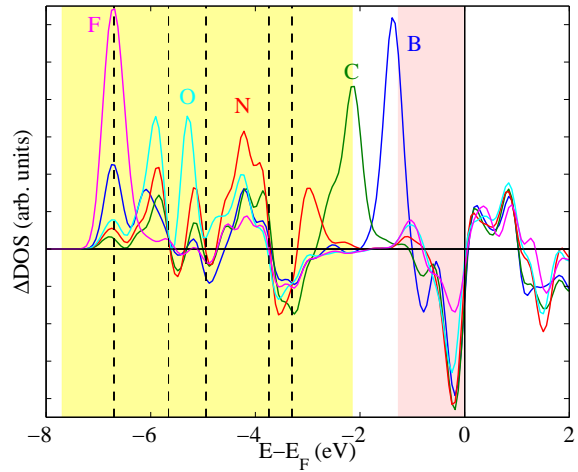}
\caption{\label{fig:VC_DeltaDOS_all_adsorbates}
  Total $\Delta$DOS for the adsorbates B, C, N, O, and F on the
  VC($111$) surface. The shaded areas and the dashed vertical lines
  represent the same quantities as in Fig.~\ref{fig:DOS_surface}.}        
\end{figure}
%

To illustrate the variations in $\Delta$DOS upon change of adsorbate,
Fig.~\ref{fig:VC_DeltaDOS_all_adsorbates} shows the calculated
$\Delta$DOS's for the period 2 adatoms on the VC($111$) surface. This 
trend has previously been studied on the TiC and TiN ($111$) surfaces
and described with the CCM.\cite{VoRuLu06,RuVoLu06,RuVoLu07,RuLu07}

For all the adsorbates there are quenchings of both the TMSR and the
CSR's, but to different degrees for different adsorbates. As we go
from left to right along the adatom period (B$\rightarrow$F), the
energy of the adatom-projected DOS decreases. At the same time, its
width changes: localized for B and C, delocalized for N and O, and
again localized for F. Also, the depletion of the TMSR decreases
gradually. The degree of depletion of the substrate CSR's varies also
between the different adsorbates. An analysis of the Kohn-Sham wave
functions shows that adatom-projected DOS peaks that lie in the lower
part of the UVB consist of strong adatom--C bonding states, while
peaks in the upper part of the UVB contain strong adatom--TM bonding
states.  

Again, these variations indicate strongly varying degrees of
interaction between adsorbate, TMSR, and CSR levels.  They resemble
the trends previously found on the TiC and TiN ($111$) surfaces and
should therefore be possible to explain in a similar way within the
CCM. This is done in Section~\ref{sec:discussion}, where the DOS
variations are related to the calculated $E_{\text{ads}}$ trends.


\section{Molecular adsorption}

%
\begin{table}
\centering
\caption{\label{tab:NHx_on_VC}
  Results from the calculations on the molecules NH$_x$ ($x=1,2,3$)
  adsorbed on the VC($111$) surface: adsorption energies
  $E_{\text{ads}}$, perpendicular distances \textit{d} between the
  molecule and the surface, and charge transfers from the surface to
  the molecule obtained by a Bader analysis.} 
\begin{tabular}{ccccccccc}
\hline
\hline
         & $E_{\text{ads}}$~(eV)  & $d$~(\AA)& Bader (units of 
$|e|$) \\ 
\hline
NH       & $5.93$                & $1.23$  & $0.96$\\
NH$_2$   & $4.63$                & $1.48$  & $0.64$\\
NH$_3$   & $0.84$                & $1.83$  & $0.14$\\
\hline
\hline
\end{tabular}
\end{table}
%

%
\begin{figure*}
\centering
\includegraphics[width=0.47\textwidth]{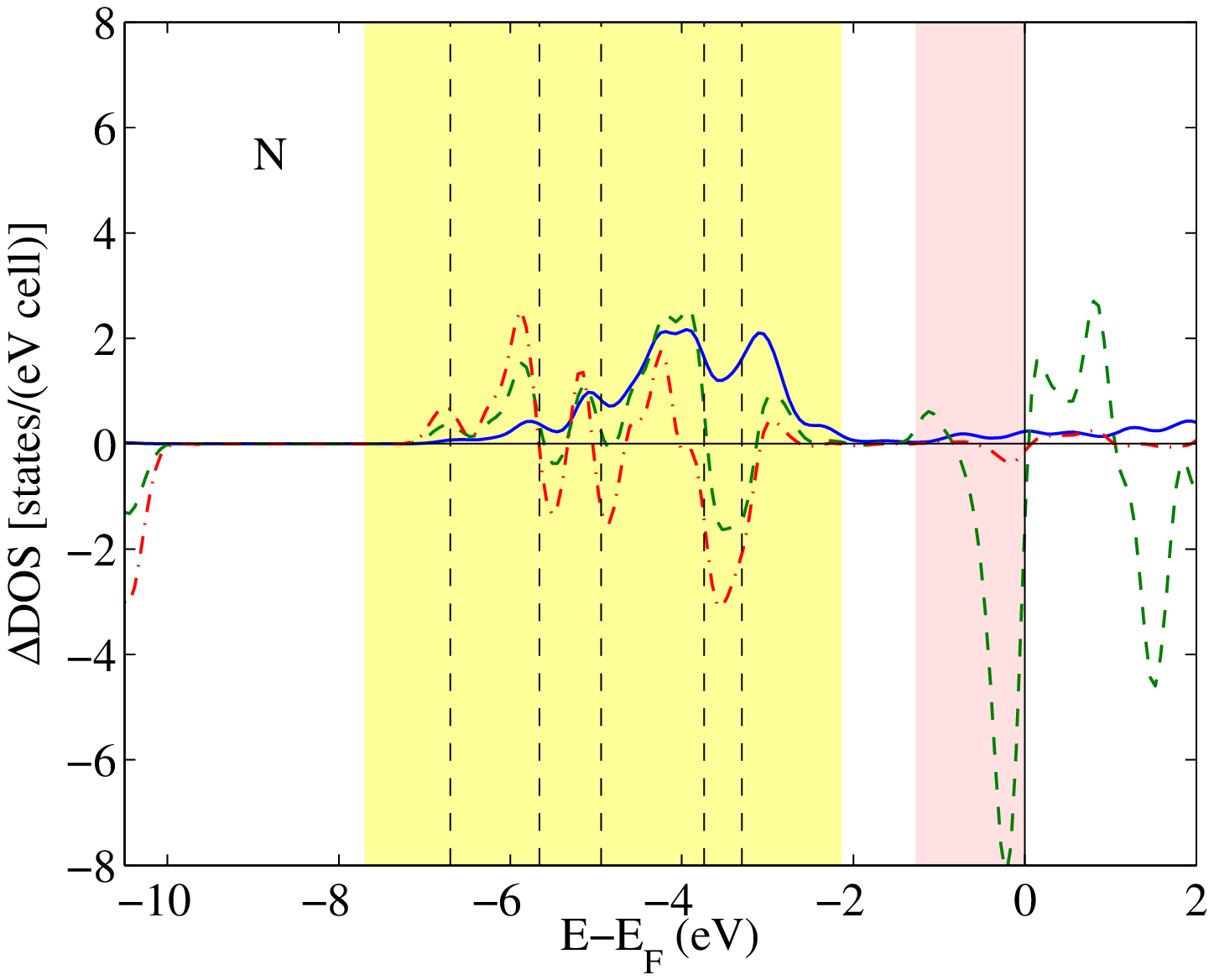}
\includegraphics[width=0.47\textwidth]{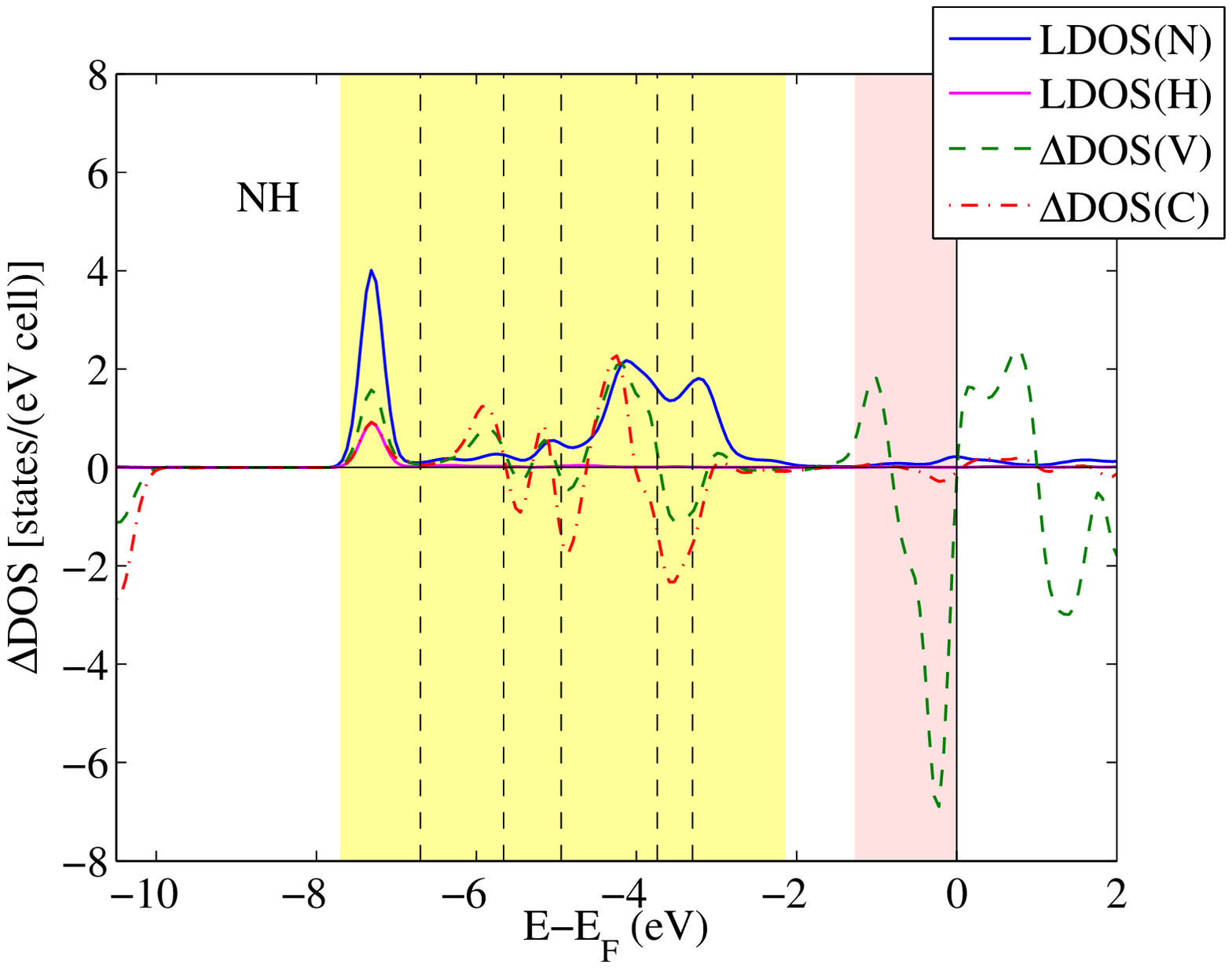}
\includegraphics[width=0.47\textwidth]{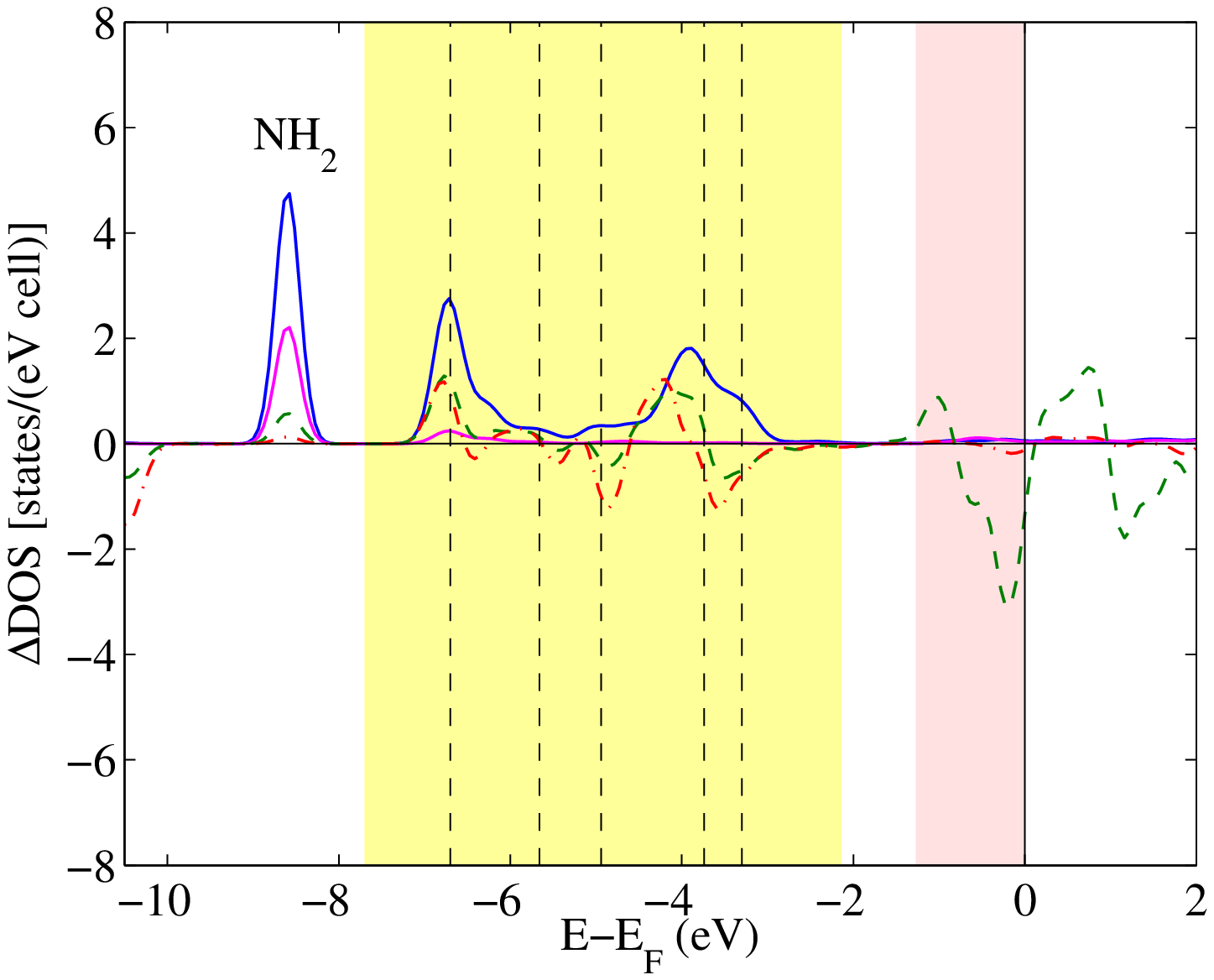}
\includegraphics[width=0.47\textwidth]{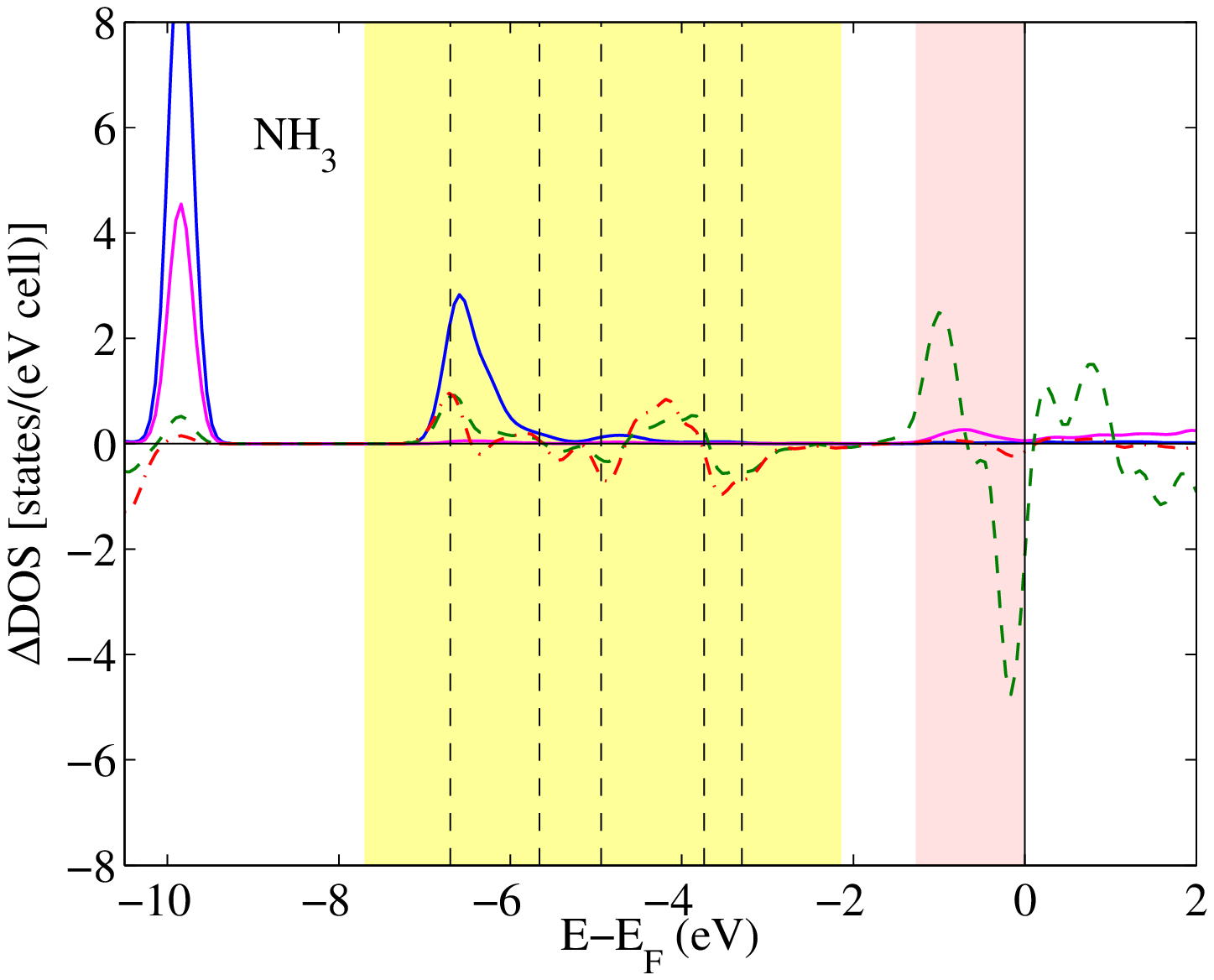}
\caption{\label{fig:VC_DeltaDOS_NHx}
  Total and atom-projected $\Delta$DOS's and LDOS's for N, NH, NH$_2$
  and NH$_3$ adsorbed on VC($111$). The shaded areas and the dashed
  vertical lines represent the same quantities as in
  Fig.~\ref{fig:DOS_surface}.}        
\end{figure*}
%

In this Section we present results for molecular adsorption on the
VC($111$) surface, which is here chosen as a prototype for molecular
adsorption on the TMC($111$) surface. The molecular adsorbates are the
NH$_x$ molecules ($x=1,2,3$), which are a part of our study in
Ref.~\onlinecite{VoHeRuLu09}. The same supercell size and
computational parameters as those used for the atomic adsorption are
employed for these calculations (see Section~\ref{sec:surf_comp}). The
molecules are adsorbed with the N atom closest to the surface in the
fcc site (the stable site for N atomic adsorption). The adsorption
energy is calculated relative to the energy of an isolated NH$_x$
molecule as  
\begin{equation}
E_{\text{ads}} = -(E_{\text{slab+NH}_x} - E_{\text{clean slab}} - E_{\text{NH}_x}).
\end{equation}
The charge transfer between substrate and absorbed molecule is
calculated as the sum of the Bader charges on each of the constituent
atoms in the molecule. 

A summary of the results is given in Table~\ref{tab:NHx_on_VC}. As the
molecule size $x$ increases, the perpendicular distance to the surface
increases, while both the adsorption energy $E_{\text{ads}}$ and the
charge transfer from surface to molecule decrease. 

Figure~\ref{fig:VC_DeltaDOS_NHx} shows calculated LDOS's (projected on
the molecule atoms) and $\Delta$DOS's (projected on the substrate
atoms) for N and NH$_x$ ($x=1,2,3$) on VC($111$). The energy and
structure of the LDOS's vary between the different adsorbates. For
atomic N, it consists of one delocalized region in the middle of the
UVB. For NH, there is a slightly weaker delocalized region of N
character in the middle of the UVB, together with a sharp peak of
mixed N and H character in the lower part of the UVB. For NH$_2$, the
delocalized region of N character in the middle of the UVB is further
weakened and a new sharp peak of mixed N and H character appears below
the UVB. For NH$_3$, the delocalized region in the middle of the UVB
disappears, while the sharp peak below the UVB is stronger and lies at
lower energy. For all adsorbates, there is a quenching of both the  
TMSR and the CSR's. These SR depletions are largest for N and decrease
every time a H atom is added to the molecule. Hence, each addition of
a H atom weakens the interaction between the adsorbed molecule and the
SR's.


\section{Discussion}\label{sec:discussion}

As described above, we find several trends in the calculated
adsorption energies: ``M''-shaped variations with respect to atomic
adsorbate, linear variations with respect to substrate, qualitatively
different types of adsorption on ScC and for F adatom, and trends in
molecular adsorption strength.   

In this Section, we describe how these trends can be described within
our previously proposed concerted-coupling model (CCM) for atomic
adsorption on TiX($111$) surfaces.\cite{RuLu07,RuVoLu06,RuVoLu07}
First, the CCM is described by giving an account of the main results
of Refs.~\onlinecite{RuLu07,RuVoLu06,RuVoLu07} and shown to apply also
on the here considered TMC's by pointing out the similarities in
electronic structure results. Then, the different adsorption-energy
trends are explained in terms of the CCM. Finally, the results lay the
ground for the single descriptor $\varepsilon_{\text{CCM}}$ for the
adsorption strength, recently introduced in
Ref.~\onlinecite{VoHeRuLu09}.

\subsection{Concerted-coupling model}

For TM surfaces, the $d$-band model yields a successful description of
electronic structure and adsorption.\cite{HammerNorskov00,Bligaard08}
For instance, its key parameter $\varepsilon_d$, the mean energy of
the substrate $d$ band, is a good descriptor for, \textit{e.g.},
adsorption.\cite{Hammer95,HammerNorskov95,BlNoLu08} Such a fact
facilitates the design of new TM catalysts by computational
screening.\cite{Abild-Pedersen07,Sehested2007,Studt,Norskov2009}   

For adsorption on TMC's, however, there are deviations from the
\textit{d}-band model.\cite{Kitchin,Fernandez} Indeed, we find that
different TMC surfaces can have the same value of $\varepsilon_d$ but
different $E_{\text{ads}}$ values. For example, the $\varepsilon_d$
values for TiC($100$) and TiC($111$) surfaces are as close as
$0.43$~eV and $0.30$~eV, whereas the $E_{\text{ads}}$ values for
oxygen differ considerably, being $5.79$~eV and $8.76$~eV,
respectively.\cite{RuLu07} 

\begin{figure*}
\centering
\includegraphics[width=\textwidth]{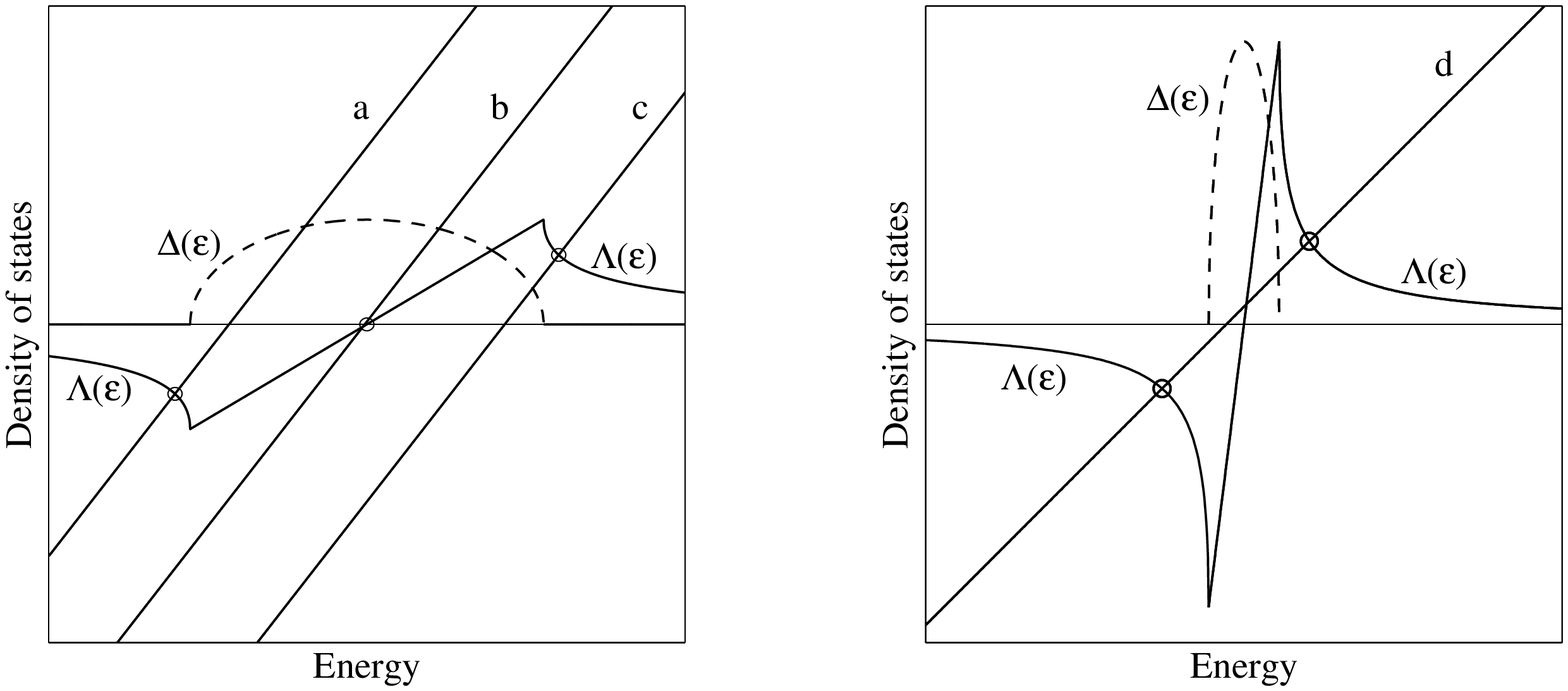}
\caption{\label{fig:NA}
  A schematic representation of the solutions of the Newns-Anderson
  model for atomic chemisorption on a metal surface.\cite{Newns} The
  left and right panels illustrate the ``weak'' and ``strong''
  chemisorption limits, respectively. The solutions, that is, the
  adatom-localized DOS's after adsorption, are yielded by the
  intersections between the straight lines a--d and the
  function $\Lambda(\varepsilon)$. The energy of the adatom frontier 
  orbital before adsorption is given by the intersection of the lines
  a--d and the $x$-axes. Thus, cases a and c correspond to a shift of
  the adlevel to lower and higher energies, respectively, while case d
  corresponds to a formation of well-separated bonding and antibonding
  levels. In case b, only a broadening of the adlevel is obtained. The
  function $\Delta(\varepsilon)$ represents the DOS of the clean
  surface before adsorption.  Adapted from Ref.~\onlinecite{Newns}.} 
\end{figure*}

In the \textit{d}-band model, adsorption trends are explained by the
interaction between the adatom frontier orbital and the narrow TM $d$
band. Using the terminology of the Newns-Anderson (NA) model for
chemisorption, \cite{Anderson,Newns,Grimley,SpDe93} this bond is
typically ``strong'' and results in the formation of separated bonding
and antibonding adatom--substrate states (see case d in
Fig.~\ref{fig:NA}).\cite{Newns} 

%
%

In Refs.~\onlinecite{RuLu07,RuVoLu06}, and \onlinecite{RuVoLu07},
atomic adsorption on the TiC and TiN ($111$) surfaces is described as
a result of two types of interactions between adsorbate and substrate.
In the terminology of the NA model, the coupling of the adatom
frontier orbital is typically ``strong'' with the substrate TiSR and
``weak'' with the substrate XSR's (X = C or N) (as in case d and in
cases a--c in Fig.~\ref{fig:NA}, respectively). A concerted action of
these couplings gives a qualitative explanation for the calculated
adsorption energy trends on the
TiC($111$)\cite{RuLu07,RuVoLu06,RuVoLu07} and
TiN($111$)\cite{VoRuLu06,RuVoLu06,RuVoLu07} surfaces. The TiSR is
present on the TiX($111$) surfaces but not on the ($001$) ones.  The
XSR's are found in the TiX($111$) UVB's.   

In the first mentioned coupling above, the large overlap of the
localized TiSR with the adatom orbital causes a strong adatom--TiSR
interaction (in the NA sense). Well-separated bonding and antibonding
states of mixed adsorbate and Ti character are then formed (case d in
Fig.~\ref{fig:NA}). The bonding-state energy lies below the
free-adatom and TiSR levels, while the antibonding state resides above
the TiSR level. 

In the other coupling, the bonding adatom--TiSR level interacts with
the XSR's present in the substrate UVB.  Due to the short range of the
XSR's (compared to the TiSR), this interaction is weak (in the NA
sense) and causes a mixture of broadening and shifting of the bonding
adlevel--TiSR state. A state located in the middle of the UVB is
mainly broadened (case b in Fig.~\ref{fig:NA}), while a state at the
edge of the UVB is mainly shifted away from the UVB center of mass
(cases a and c in Fig.~\ref{fig:NA}).  

Evidence for this concerted coupling is given by detailed analyses of
the calculated adsorbate-induced DOS's ($\Delta$DOS's) and of
real-space visualizations of the Kohn-Sham wave functions.  These show
that upon adsorption, there are 
(i) a sharp decrease in DOS at the TiSR energy;  
(ii) a sharp increase of DOS just above the TiSR energy; 
(iii) depending on the adatom species, narrow or broad bands of mainly
adatom character at the edge of or within the substrate UVB energy
range, respectively;   
(iv) a depletion of X-localized UVB states at the XSR energies; and 
(v) a formation of adatom-localized sub-peaks in between the energies
of the substrate XSR's. Points (i)--(iii) show that bonding and
antibonding adlevel--TiSR states are formed, while points (iii)--(v)
prove a coupling between the bonding adlevel--TiSR state and the
XSR's in the substrate UVB.   

More detailed analyses of the trends in calculated $\Delta$DOS's and
Kohn-Sham wave functions for second- and third-period adatoms on
TiX(111) surfaces show that (i) the magnitude of the DOS reduction at
the TiSR energy decreases successively as the adatom number $Z$
increases along a period and (ii) the adatom--X bonding character of
the adatom-localized peaks increases as the adatom number $Z$
increases along a period.  These trends, which arise from the
successive lowering of the adlevel energy as $Z$ increases, indicate
that the adatom--TiSR coupling decreases in strength along each adatom
period, towards the right in the periodic table, while the
contribution of the XSR's to the bonding increases.   

Such trends provide a basis for understanding the calculated trends in
atomic adsorption energies. For instance, the maximum for the group-VI
adatoms (O and S) is explained in terms of a stronger coupling to the
substrate CSR's.  On the other hand, the weaker bonding of group-VII
adatoms (F and Cl) is explained by a weakened coupling to the CSR's,
due to the adatom state being almost fully ionized by its interaction
with the TiSR. Thus, the F and Cl adsorbates lack almost any covalent
interaction with the substrate UVB and their adsorption is practically
ionic in nature, as confirmed by Bader and charge-density
analyses. Also, the almost equal adsorption strengths of C and N can
be explained in terms of the opposite trends in adatom--TiSR and
adatom--CSR coupling strengths found for varying $Z$ values.   

On the same basis, the CCM is also able to describe changes in
adsorption energies arising from changes in adsorption site, {\it
  i.e.}, between fcc, hcp, and top O adatoms on
TiC($111$).\cite{RuLu07}

\subsection{Generalization of the CCM to other TMC's}

The calculated energetics and electronic structures of atomic
adsorption on the TMC($111$) surfaces described in
Section~\ref{sec:ads} can be analyzed in the same way as described
above for TiX($111$).    

The fact that upon adsorption the TMSR's and CSR's are quenched on all
considered surfaces shows that both types of resonance participate in
the substrate--adatom bond. The strong character of the adatom--TMSR
bonds is evidenced by the formation of well-separated bonding and
antibonding states of mixed adatom and TM character, at energies below
and above the TMSR's, respectively. The weak character of the
adatom--CSR bonds is confirmed by the broadening and/or shift of the
bonding adatom--TMSR level.   

The exception is ScC($111$), on which the adatom--CSR coupling
approaches a strong NA character, with a strong depletion of CSR
states in the middle of the UVB and the formation of separated bonding
and antibonding states below and above the UVB, respectively.  This is
due to the stronger CSR's on this surface, which in turn arises from
the stronger C--C bonds in bulk ScC than in the other
TMC's.\cite{VoRu09} Also, the adatom--TMSR coupling is qualitatively
different on ScC(111), compared to the other TMC's, as the TMSR of
clean ScC(111) lies above $E_F$ and is thus empty before adsorption.

Thus, the basic adsorption mechanism is the same on all the considered
TMC($111$) surfaces, implying that the general applicability of the
CCM includes TMC($111$) surfaces.  In addition, ScC(111) provides a
playground for testing the contribution of the adlevel--CSR coupling
to the total adsorption strength.   

In the following, we give more explicit support to the applicability
of the CCM to the TMC($111$) surfaces by showing that the CCM can be
used to understand the major features of the calculated atomic and
molecular adsorption-energy trends on the considered TMC($111$)
surfaces.

\subsection{Adsorption trends with respect to substrate and CCM}
\label{sec:disc_substrate}

Figure~\ref{fig:Ads_on_TMX_all_ads_sub}(b) shows the calculated energy
values for atomic adsorption on TMC($111$) surfaces as functions of
the substrate TM group number.  In this subsection we show that the
CCM can capture its main trends. 

In a simple molecular two-level picture, the strength of a bond is
related to the energy shift between the levels of the isolated
constituents and of the hybridized states resulting from the coupling.
In a simplified version of the Newns model, an adsorbate and the
active part of the substrate can be viewed as such a
``molecule''.\cite{Newns}  Accordingly, from the calculated
$\Delta$DOS's the bond strength can be extracted as the energy
difference between the final bonding adatom level (\textit{i.e.},
after coupling to both TMSR and CSR's) and the TMSR peak in the
surface DOS. The CCM is then able to explain the trends in bond
strength by relating this energy difference to concerted adatom--TMSR
and adatom--CSR interactions.   

In the following, we employ such an approach to understand the
qualitatively different trends in $E_{\text{ads}}$ values for C and O
adatoms, as the substrate TM group number increases from IV to VI
[Fig.~\ref{fig:Ads_on_TMX_all_ads_sub}(b)].   

For C, the $E_{\text{ads}}$ values are approximately constant between
groups IV and VI.  The calculated $\Delta$DOS's (see
Fig.~\ref{fig:C_DeltaDOS_all_substrates}) show that the final bonding
adatom level lies at the upper edge of the UVB on all considered
substrates. Thus on all these substrates, the energy difference
between the final adatom level and the original substrate TMSR is
approximately constant. This implies that the result of the concerted
action of the adatom--TMSR and adatom--CSR's interactions is
approximately constant.  In particular,
Fig.~\ref{fig:C_DeltaDOS_all_substrates} shows that the DOS's of all
substrates resemble the same NA type of adlevel--CSR's interaction,
that is, type c in Fig.~\ref{fig:NA}.  

For O, with its lower $2p$ level, the $E_{\text{ads}}$ values decrease
strongly as the substrate TM group number increases from IV to
VI. This is reflected in the calculated $\Delta$DOS's
(Fig.~\ref{fig:O_DeltaDOS_all_substrates}) by a strongly decreasing
energy difference between the final adatom level and the original
substrate TMSR peak. Thus, in contrast to C, for O the concerted
action of the two types of coupling yields qualitatively different
results.  In particular, Fig.~\ref{fig:C_DeltaDOS_all_substrates}
shows DOS's that correspond to adlevel--CSR's interactions that vary
in a qualitative way from (\textit{cf.\ }Fig.~\ref{fig:NA}) NA case a
(on group IV), to NA case b (on group V), and to NA case c (on group
VI). As a consequence, the adlevel--CSR's interactions cause a
downward energy shift (on group IV), a broadening (on group V), and an
upward energy shift (on group VI) of the bonding adlevel.   

We also notice that for both adsorbates the antibonding adatom--TMSR
level lies above $E_F$ (and is thus empty) for all substrates except
the group-VI TMC's $\delta$-MoC and WC.  This filling of antibonding
states implies an extra decrease in adatom--TMC bond strength, as
group V $\rightarrow$ VI. Indeed, this can be seen in the calculated
$E_{\text{ads}}$ trend for C
[Fig.~\ref{fig:Ads_on_TMX_all_ads_sub}(b)].   

For all adsorbates except F, the calculated $E_{\text{ads}}$ values
increase from ScC($111$) to TiC($111$). As described above, the
adlevel--CSR's interaction is more pronounced on ScC($111$) than on
the other TMC's, thus approaching the strong NA limit. Also, the
adatom--TMSR coupling is qualitatively different on ScC($111$), due to
the TMSR of clean ScC($111$) lying above $E_F$ and being thus empty
before adsorption. The weaker adsorption on ScC($111$) can thus be
understood to be due to the qualitatively different interaction with
the two types of SR's.

\subsection{Adsorption trends with respect to adsorbate and CCM}

The adsorption-energy trends with respect to adsorbate
[Fig.~\ref{fig:Ads_on_TMX_all_ads_sub}(a)] show ''M``-shaped
$E_{\text{ads}}$ trends for most of the considered TMC's.  The trend
on TiC($111$) was analyzed in Ref.~\onlinecite{RuLu07} and explained
within the CCM to arise from competing opposite trends in
adsorbate--TMSR and adsorbate--CSR's interactions.   

The calculated $\Delta$DOS's for the here considered TMC's
(illustrated for VC in Fig.~\ref{fig:VC_DeltaDOS_all_adsorbates}) show
trends that are similar to those found for TiC. At the same time, the
$E_{\text{ads}}$ trends for all considered TMC's resemble the one on
TiC($111$), with a local maximum for the O adatom. However, on group-V
TMC's the calculated $E_{\text{ads}}$ values for O are approximately
equal to the ones for C, and on group-VI TMC's the O values are
smaller than the ones for C. Thus, the chemisorption strength of O
appears to weaken relative to C as the substrate TM group number
increases.  This can be understood to be a consequence of the
different $E_{\text{ads}}$ trends for C and O that were described in
Sec.~\ref{sec:disc_substrate} above, with the adsorption strength
decreasing much faster for O than for C as the TM group number
increases from IV to VI.   

On the other hand, the $E_{\text{ads}}$ values for ScC($111$) show a
monotonic increase between B and O adatoms. As described above, the
adlevel--CSR's interaction is stronger on ScC($111$) than on the other
considered TMC($111$) surfaces. Therefore, the monotonically
increasing trend (as B $\rightarrow$ O) of the adlevel--CSR's coupling
contribution to the adatom--TMC bond is stronger on ScC($111$).  

Finally, on all substrates there is a sharp decrease in adsorption
energy, as O $\rightarrow$ F.  This arises from the weakened
adlevel--CSR's interaction, which is due to the fact that the adatom
is almost fully ionic from its interaction with the TMSR, as described
previously on TiC($111$).\cite{RuLu07}

\subsection{Molecular adsorption and CCM}

The adsorption energies of the considered molecular adsorbates NH$_x$
($x = 1,\ 2,\ 3$) on VC($111$) show a decreasing trend as $x$
increases (Table~\ref{tab:NHx_on_VC}). This trend can be understood in
terms of the CCM as arising from the concerted coupling between each
molecular orbital and the two types of SR's.  

The quenchings of both TMSR and CSR's in the calculated $\Delta$DOS's
for NH$_x$ on VC($111$) (Fig.~\ref{fig:VC_DeltaDOS_NHx}) indicate the
presence of couplings to both types of SR's. In addition, the
adsorbate-projected LDOS's show strongly bound sharp peaks of mixed N
and H character as well as more delocalized regions of only N
character within the UVB.  A comparison with calculated DOS's for the
free NH$_x$ molecules shows that the sharp peaks correspond to
low-energy molecular levels that interact weakly with the substrate
SR's.  On the other hand, the delocalized regions arise from the
coupling of higher-energy molecular levels with the SR's, in a way
that is similar to the adsorption of atomic
N. Figure~\ref{fig:VC_DeltaDOS_NHx} shows that as the number $x$ of H
atoms in the molecule increases, the interaction of the higher-energy
molecular levels with the SR's weakens, thus lowering the adsorption
strength.

\subsection{Applicability of the CCM}

The above discussion shows that for the atomic and molecular adsorption on the 
TMC($111$) surfaces a picture based on the concerted action of two types of 
adatom--substrate interactions applies.  It can be used to describe key 
features of the calculated adsorption-strength and electronic-structure trends.  
In this concerted-coupling model (CCM), two types of SR's participate in the bond, 
TM-localized SR's (TMSR) and C-localized SR's (CSR).  Therefore we conclude that 
the CCM is valid for atomic adsorption on all the considered TMC(111) surfaces.

\subsection{Descriptor for adsorption on TMC($111$): $\varepsilon_{\text{CCM}}$}

To be able to describe the variations in adsorption strength in a
simple yet efficient way, a descriptor $\varepsilon_{\text{CCM}}$,
defined as the mean energy of the TMSR, can be
introduced.\cite{VoHeRuLu09}  Due to the approximately constant energy
difference between TMSR and CSR's in the considered compounds, such a
descriptor is able to capture the important variations of both TMSR
and CSR energy trends while being at the same time both conceptually
simple and measurable or calculable.   

\begin{figure}
\centering
\includegraphics[width=0.5\textwidth]{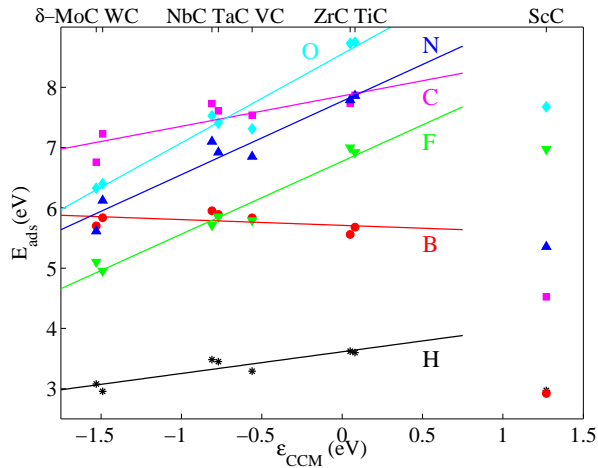}
\caption{\label{fig:E_ads_vs_epsilonCCM}
  Correlation between $E_{\text{ads}}$ and $\varepsilon_{\text{CCM}}$
  for the atomic adsorbates H, B, C, N, O and F.} 
\end{figure}

This is confirmed by Fig.~\ref{fig:E_ads_vs_epsilonCCM}, which shows a
linear correlation between our calculated $E_{\text{ads}}$ values and
$\varepsilon_{\text{CCM}}$ for each of the considered atomic
adsorbates, also when the results of Ref.~\onlinecite{VoHeRuLu09} are
significantly extended. The exception is ScC($111$), which is a
consequence of the mentioned qualitative difference in electronic
structure of this surface, that is, an empty TMSR and a strong
adlevel--CSR coupling.  As discussed above, the variations in the
gradients of the lines for the different adsorbates can be understood
within the CCM from the details in the interactions between the
adlevel and the different SR's.   

In Ref.~\onlinecite{VoHeRuLu09}, linear correlations are also found
between $\varepsilon_{\text{CCM}}$ and the molecular adsorption
energies of NH, NH$_2$, and NH$_3$, as well as the activation-energy
barriers for N$_2$.  This is a consequence of the fact that both
atomic and molecular adsorptions appear to follow the same basic
mechanisms of the CCM, as argued above.  As a consequence, we show in
Ref.~\onlinecite{VoHeRuLu09} that scaling and
Br{\o}nsted-Evans-Polanyi (BEP) relations apply for adsorption on
TMC(111) surfaces. Such findings are of importance for the design of
new catalysts.


\section{Conclusions}\label{sec:conclusion}

The possibility to understand materials has today increased
considerably, as DFT has developed into a predictive theory. One
obvious application of DFT is to calculate numbers for bonding
strengths, structure parameters and coordinates, \textit{etc.}
Another, slightly more demanding, application is to look for more
detailed information and develop conceptual frameworks and models in
which we can formulate our understanding and on which we can base our
further thinking, including ideas about new materials. We have found
that various densities of states (DOS's) are excellent tools for such
understanding and vehicles for thought. In our approach, extensive use
of total, local, projected, and difference DOS's are cornerstones in
the building-up of a conceptual framework and model.   

This study deals with understanding the atomic and molecular
adsorption on the TMC($111$) surfaces using such a detailed electronic
structure analysis approach. By extensive DFT calculations on ScC,
TiC, VC, ZrC, NbC, $\delta$-MoC, TaC, and WC, we study trends in clean
surface properties, adsorption energies, and various DOS's.  In brief,
we find that only a certain part of the surface-localized electronic
structure, corresponding to the TM-localized (TMSR) and the
C-localized (CSR) surface resonances (SR's) are of importance to
understand the trends in adsorption energies from one TMC surface to
another and from one adsorbate to another. Despite the TM termination
of the investigated TMC surfaces, the second-layer C atoms are found
to play a crucial role in the chemisorption. This is particularly
evident on the ScC($111$) surface, where the TMSR is empty and where
the CSR's are particularly strong.  

Having thus identified the key parameters, we possess the foundations
for a concerted-coupling model (CCM) in which trends in adsorption
strength are the result of a concerted action of both adsorbate--TMSR
and adsorbate--CSR's couplings. This has earlier been shown for the
TiC and TiN ($111$) surfaces\cite{RuLu07,RuVoLu06,RuVoLu07} but the
breadth and versatility are here shown by applications to 
other TMC's and trends, including: adsorption trends with respect to
substrate, adsorption trends with respect to atomic adsorbate,
adsorption trends on ScC($111$), adsorption trend for group-VII adatoms,
and molecular-adsorption trends. This deepens the interpretation of
the model and broadens its usefulness. 

The applicability of the CCM to both atomic and molecular adsorption
opens up the possibility to study reactions for, \textit{e.g.},
catalytic applications on this class of materials. For example, it
allows the formulation of a single descriptor
$\varepsilon_{\textrm{CCM}}$ for the adsorption
strength.\cite{VoHeRuLu09} Also, it implies the existence of scaling
relations between molecular and atomic adsorption strengths as well as
Br{\o}nsted-Evans-Polanyi relations.\cite{VoHeRuLu09} 

Since the CCM framework is based on rudimentary bonding principles
with general applicability, we believe that it is possible to
generalize it to other materials that possess surface-localized
states. We have already shown its applicability to atomic adsorption
on TiN($111$)\cite{RuVoLu06,RuVoLu07} and believe that the same
chemisorption mechanism should be valid for other nitrides. Ligand and
vacancy systems have been shown to belong to the group of materials
where the CCM applies,\cite{VoHeRuLu09} as do certain TM
surfaces,\cite{VoHeRuLu09} where, however, it does not need to replace
the sufficient and natural \textit{d}-band model. Natural extensions
should include TM oxides, sulfides, and borides. Design of materials,
including atomic-scale engineering, is also an enticing prospect for
further applications.


\section*{Acknowledgments}
Valuable discussions with Anders Hellman are acknowledged. The
calculations were performed at HPC2N and NSC via the Swedish National
Infrastructure for Computing. B.~I. Lundqvist gratefully acknowledges
support from the Lundbeck foundation (Denmark) via the Center for
Atomic-scale Materials Design.


\end{document}